\documentclass[pra,amssymb,amsmath,showpacs]{revtex4}

\usepackage{graphicx}
\usepackage{amssymb}
\usepackage{amsmath}
\usepackage{color}

 \usepackage[varg]{txfonts}	% fonts for integrals etc

\def\url#1{\textcolor{blue}{\underline{#1}}}	% web references
\definecolor{oneblue}{rgb}{0,0,0.65}
\definecolor{onered}{rgb}{0.65,0,0}
\definecolor{turquoise}{rgb}{0,0.4,0.5}

\def\url#1{\textcolor{blue}{\underline{#1}}}	% web references
\def\note#1{\textcolor{blue}{#1}}    		% notes in blue
\def\b#1{\textcolor{oneblue}{#1}} 
\def\newb#1{\textcolor{oneblue}{#1}}		% corrected math

 \large\normalsize	 %double space

\begin{document}

% ------------------------------------------------------------
% Title page
% ---------------------------------------------------------

\title{A generalized theory for current-source density analysis in
brain tissue}

\author{Claude B\'edard and Alain Destexhe\footnote{Corresponding
author.  Tel: +33 1 69 82 34 35, Fax: +33 1 69 82 34 35, Email:
Destexhe@unic.cnrs-gif.fr}}

\affiliation{Unit\'e de Neurosciences, Information et Complexit\'e
(UNIC), \\ CNRS, \\ 1 Avenue de la Terrasse (Bat 33), \\ 91198
Gif-sur-Yvette, France \\ \ \\ {\it Physical Review E}, in press,
2011.}

\date{\today}

% ------------------------------------------------------------
% Abstract
% ---------------------------------------------------------

\begin{abstract}

The current-source density (CSD) analysis is a widely used method
in brain electrophysiology, but this method rests on a series of
assumptions, namely that {the surrounding extracellular medium is
resistive and uniform, and in some versions of the theory, that the
current sources are exclusively made by dipoles.}  Because of these
assumptions, this standard model {does not correctly describe}
the contributions of monopolar sources or of non-resistive aspects
of the extracellular medium.  We propose here a general framework
to model electric fields and potentials resulting from current
source densities, without relying on {the above} assumptions. 
We develop a mean-field formalism which is a generalization of the
standard model, and which can directly incorporate non-resistive
(non-ohmic) properties of the extracellular medium, such as ionic
diffusion effects.  This formalism recovers the classic results of
the standard model such as the CSD analysis, but in addition, we
provide expressions to generalize the CSD approach to situations
with non-resistive media and arbitrarily complex multipolar
configurations of current sources.  We found that the power
spectrum of the signal contains the signature of the nature of
current sources and extracellular medium, which provides a direct
way to estimate those properties from experimental data, and in
particular, estimate the possible contribution of electric
monopoles.

\end{abstract}


\maketitle

% ------------------------------------------------------------
% Text
% --------------------------------------------------------
%\clearpage
\section{Introduction}

The current-source density (CSD) analysis \cite{Mitzdorf,Pettersen}
is a method consisting of estimating the underlying current sources
from a series of recordings of the extracellular electric
potential.  This method is widely used in neuroscience, and applies
well to layered structures of the brain, such as cerebral cortex,
hippocampus or cerebellum \cite{Mitzdorf}.  The CSD analysis is
based on the ``standard'' model of electric potentials in
biological tissue \cite{Mitzdorf,Pettersen,Plonsey}, which rests on
{the hypothesis that the extracellular medium is resistive
(ohmic) and uniform.}  Other influences, such as ionic diffusion,
are assumed to play a negligible role on the propagation of the
electric field.  

% {Moreover, in some versions of the CSD
% theory~\cite{Mitzdorf,Linden2010}, the sources of the electric field
% are assumed to be exclusively made by dipoles.} 

Based on this set of hypotheses, the equation that determines the
electric potential at macroscopic scales (\b{$\sim 50~\mu m$} or
more) is given by: 
\b{
\begin{equation}
 \nabla\cdot(\sigma^e\nabla V)=\sigma^e\nabla^2 V=\frac{\partial\rho}{\partial t} ~ ,
\label{standard}
\end{equation}}
where \b{$\sigma^e$} is the electric conductivity of the
extracellular medium.  This expression can be obtained by applying
the differential law of charge conservation and Ohm's law.  The
term \b{$-\frac{\partial\rho}{\partial t}$} is interpreted as the
{volumic density \b{$I_m$}} of current sources.  This equation
forms the basis of the CSD analysis method
\cite{Mitzdorf,Pitts,Pettersen,Nicholson1975}.  

According to Eq~\ref{standard}, the electric potential $V$ would
only depend on electric conductivity and not at all on electric
permittivity.  However, Poisson's law {in a homogeneous
medium} \b{$(\varepsilon\nabla^2 V= -\rho)$} implies that \b{V}
will be twice smaller for twice larger \b{$\varepsilon$} with the
same charge distribution, so it is paradoxical that permittivity is
not taken into account in CSD analysis.  Moreover, according to
Eq~\ref{standard}, the electric potential is determined solely by
the charge conservation law, and independently of Poisson's law,
which is contradictory with Gauss' law in Maxwell equations.  

If we take Gauss' law into account, we can write
\b{
$$
 \frac{\partial\rho}{\partial t} = \sigma^e\nabla^2 V
 = -\frac{\sigma^e}{\varepsilon}\rho,
$$}
The general expression for \b{$\rho$} is:
\b{
$$
\rho(\vec{x},t) = \rho(\vec{x},0)e^{-\frac{t}{\tau_{\mbox{\tiny MW}}}}
$$}
where \b{$\tau_{\mbox{\tiny MW}}=\frac{\varepsilon}{\sigma^e}$} is
the Maxwell-Wagner time of the medium. However,
\b{$\tau_{\mbox{\tiny MW}}$} is usually considered as negligible
(typical values for biological tissue are \b{$\tau_{\mbox{\tiny
MW}}\approx 10^{-10}~s$}, with \b{$\sigma=0.3~S/m$} and
\b{$\varepsilon \approx 10^{-10}~F/m$}) such that the current
source density must be approximately zero, which is paradoxical. 
One way to resolve this paradox is to consider that the
Maxwell-Wagner time is not negligible, or that electric parameters
display strong spatial variations.  However, such conditions
contradict the hypothesis that the medium is resistive, and lay
outside the domain of validity of Eq.~\ref{standard} because in
this case the impedance of the extracellular medium is complex in
Fourier space~\cite{Bed2004,BedDes2009a}.

Despite this paradox, the standard model seems to apply relatively
well to media such as brain tissue \cite{Mitzdorf,Pettersen}.  This
model has, however, the drawback that it cannot be used to
determine the validity of the hypotheses it is based on.  Moreover,
there is no clear definition of microscopic or macroscopic levels,
and consequently, it is difficult to include possible frequency
dependencies that could result from different physical phenomena at
intermediate (mesoscopic) scales, such as ionic diffusion or
membrane polarization \cite{BedDes2009a}.

{In the present paper, we introduce a more general formalism
which does not rest on the hypotheses of classic CSD where the
medium is hypothesized to be uniform and resistive, which also
supposes that the electric parameters are constant and independent
of frequency.}  The goal of this new formalism is to provide
generalized expressions for CSD analysis in non-resistive media. 
Our aim is also to provide a theory which is general enough to
enable testing different hypotheses concerning the nature of
current sources and the electrical properties of the surrounding
extracellular medium, which could then be directly estimated from
experimental data.

\section{General Theory}

In this section, we derive a mean-field theory of the electric
field and potential resulting from current densities in biological
tissue, by staying as general as possible.

\subsection{Definitions and scales}

{In the generalized formalism presented below, we will define
the current sources from conductance variations.  We will assume
that the differential law of charge conservation holds in a given
domain \b{D}, without defining a current source density per unit
volume.  We assume that the conductance variations in cellular
membranes (especially around synapses), {within domain
\b{D}}, are the principal origin of the extracellular electric
field.}  
% By applying the differential law of charge conservation, we
% obtain:}
% \b{
% $$
% I=\iint\limits_{\partial D}\vec{\j}\cdot\hat{n}~dS = 
% \iiint\limits_{ D}\nabla\cdot\vec{\j}~~dv 
% =-\iiint\limits_{ D}\frac{\partial\rho}{\partial t}^{free}~dv
% $$}
{This assumption is more realistic and biological compared to
the ``classic'' assumption which is based on current source
densities per unit volume because real current sources are caused
by the opening or closing of membrane conductances in neurons.}
{Note that these two different points of view can be
complementary if we assume that the volumic density of current
source \b{$I_m$} is \b{$$I_m =- \frac{\partial\rho}{\partial t} ~ .
$$} In this case, the two points of view are mathematically
equivalent.  This is the reason why we wrote the source term as
\b{$- \frac{\partial\rho}{\partial t}$} in Eq.~\ref{standard}.}

{An important assumption of the present formalism is that all
observable phenomena can be modeled by fields which are twice
differentiable (class \b{$C^2$}).  While most fields will obey this
criterion, it will considerably simplify the mathematical analysis
to its simplest expression (commuting spatial and temporal
first-order derivatives).  In mean-field physics, by virtue of the
Stone-Weierstrass theorem, it is always possible to make a
uniform-convergence approximation of the observable phenomena by a
mean-field model of class \b{$C^2$}.  Indeed, because the
mean-field of a discontinuous field of first kind is necessarily a
continuous field (the primitive of a discontinuous function of
first kind is continuous), this restriction to class \b{$C^2$}
fields will not limit the applications of the theory developed
here.  Moreover, the theory will remain general because most fields
are class \b{$C^2$} in practice.  It will not apply to very
particular models, such as fields involving surfaces with
infinitely small thickness, current sources without volume, fields
that necessitate infinite energies\footnote{{In such
particular cases, the partial and first-order derivatives of fields
\b{$\vec{E}$}, \b{$\vec{D}$}, \b{$\vec{B}$} and \b{$\vec{H}$} are
not defined for every point of space and time.}}}

{Finally, the formalism developed below is only valid for
well-defined ranges of spatial and temporal scales.  We will
consider scales greater than \b{$1~\mu m$} (about 300 times the
size of ions such as \b{$K^+$} or \b{$Na^+$}, including
solvatation). This scale is chosen large enough for classic
electromagnetism theory to apply without ambiguity (although in
principle it can apply to scales down to a few nm).  We will define
as ``microscopic'', scales of the order of 1~$\mu$m, while
``macroscopic'' scales will be of the order of 50~$\mu$m or more. 
We will also consider the typical range of frequencies of
electrophysiological signals, up to \b{$10~kHz$}, for which the
quasistatic approximation is valid.} \\

\subsection{Mean-field Maxwell theory} \label{MF}

We start from Maxwell equations where we consider the spatial
averages of the fields and electric parameters, which will be
denoted here by brackets $< ... >$.  The spatial average is made
over some reference volume, which is invariant.  Because of the
scale invariance of Maxwell equations (e.g., see
\cite{Jackson1962}, chapter 4), the spatial averages of electric
field \b{$\vec{E}$}, electric displacement \b{$\vec{D}$}, magnetic
induction \b{$\vec{B}$} and magnetic field \b{$\vec{H}$} are linked
by the following linear operatorial equations:
\b{
\begin{equation}
 \begin{array}{rclcccrcl}
  \nabla\cdot<\vec{D}> &=& <\rho^{free}>
&~&~&~& \nabla\cdot<\vec{B}> &=& 0 \\\\
  \nabla\times<\vec{E}> &=& -\frac{\partial <\vec{B}>}{\partial t} 
&~&~&~& \nabla\times<\vec{H}> &=& <\vec{\j}>+\frac{\partial<\vec{D}>}{\partial t}
 \end{array}
\label{Mwell}
\end{equation}}
where \b{$<\vec{\j}>$} and \b{$<\rho^{free}>$} are the spatial averages of
the current density and free charge density, respectively. 

These equations allow one to find the general regularities that all
models must satisfy.  For example, the laws of energy conservation
and momentum conservation can be deduced from this set of equations
\cite{Jackson1962}.  In particular, by using the relation
\b{$\nabla\cdot(\nabla\times\vec{C})=0$} (which is in general true
for all vectorial fields of class \b{$C^2$}), one can deduce the
differential law of charge conservation:
\b{
\begin{equation}
 \nabla\cdot(\nabla\times<\vec{H}>)=\nabla\cdot<\vec{\j}>+\nabla\cdot\frac{\partial<\vec{D}>}{\partial t}=\nabla\cdot<\vec{\j}>+\frac{\partial \nabla\cdot<\vec{D}>}{\partial t}=
\nabla\cdot<\vec{\j}>+\frac{\partial <\rho^{free}>}{\partial t}=0
\end{equation}
}

However, because the above equations relate the spatial averages of
interaction fields (\b{$\vec{E},\vec{D}$},\b{$\vec{B},\vec{H}$})
with the spatial averages of the two matter fields
(\b{$\vec{\j},\rho^{free}$}), it is necessary to complete them with a
specific physical model to apply them to a given biological medium.
This specific model must allow measuring spatial averages at a
scale which is determined by the measurement method (type of
electrode for example).  Thus, the measurement system determines a
minimal reference volume, which necessarily implies to use a
mean-field formalism.

In general, (for all media of class \b{$C^2$}), the fields
\b{$<\vec{E}>$}, \b{$<\vec{D}>$, $<\vec{H}>$} and \b{$<\vec{B}>$}
are linked by the following general equations:
\b{
\begin{equation}
\begin{array}{rclcc}
<\vec{D}^*>(\vec{r},t) &=& \int_{-\infty}^{+\infty}<\varepsilon>(\vec{r},\tau,\vec{E},\vec{H})~<\vec{E}>(\vec{r},t-\tau)~d\tau ~+ <\vec{C}>(\vec{r},t,\vec{E},\vec{H}) \\ \\
 <\vec{B}>(\vec{r},t) &=& \int_{-\infty}^{+\infty}<\mu>(\vec{r},\tau,\vec{E},\vec{H})~<\vec{H}>(\vec{r},t-\tau)~d\tau ~
\end{array}
\label{lien1}
\end{equation}}
where \b{$\mu$} and \b{$\varepsilon$} are respectively the absolute
magnetic permeability and absolute electric permittivity tensors. 
Here, we have defined \b{$<\vec{D}^*>~= ~  <\vec{D}>+ <\vec{C}>$}
where \b{$<\vec{C}>$} is the source field resulting from
conductance variations.  Note that in classic electromagnetism, one
defines the electric displacement relative to vacuum permittivity
\b{$\varepsilon_{\infty}$} by
\b{$<\vec{D}_{\omega}>~=~\varepsilon_{\infty}<\vec{E}_{\omega}> +
<\vec{P}_{\omega}>$} (in frequency space)\footnote{\b{$\vec{E}$} is
the effective electric field and the polarization field
\b{$\vec{P}$} is produced by polarization of molecules and cell
surface polarization.  In general, the relation between these
vectors is algebraic in Fourier space, and thus a convolution
integral in temporal space.}, which expresses the fact that the
polarization field is proportional to the electric field through
electric susceptibility \b{$<\chi_{\omega}>$}
(\b{$<\vec{P}_{\omega}>~=~<\chi_{\omega}><\vec{E}_{\omega}>$}). 
It follows that
\b{$<\vec{D}_{\omega}^*>~=~<\varepsilon_{\infty}><\vec{E}_{\omega}> +
<\vec{P}_{\omega}>~+~<\vec{C}_{\omega}>=~ <\varepsilon_{\omega}><\vec{E}_{\omega}> +
<\vec{C}_{\omega}>$}. 

Considering Maxwell-Gauss' law
(\b{$\nabla\cdot<\vec{D}>~=~<\rho^{free}>$}) and the definitions of
the interaction fields
imply the following relations between charge density and
interaction fields: 
\b{
\begin{equation}
\left \{
\begin{array}{rcl}
\nabla\cdot<\vec{D}^*>&=&<\rho_e^{free}>\\\\
\nabla\cdot<\varepsilon_{\infty}><\vec{E}>&=&
<\rho_e^{free}>+<\rho^{\Delta~cond}>+<\rho^{bound}>\\\\
 \nabla\cdot<\vec{P}>&=&-<\rho^{bound}>\\\\ 
\nabla\cdot<\vec{C}>&=&
- <\rho^{\Delta~cond}>
\end{array}
\right .
\label{deplac}
\end{equation}}
where \b{$<\rho^{\Delta~cond}>$} represents the average variation
of free charge density produced by conductance variations, and
\b{$<\rho_e^{free}>$} is the average free charge density which does
not result from membrane conductance variations.  Note that the
divergence of the field \b{$<\vec{C}>$} depends on the exact
mechanism of conductance variation.  If this mechanism does not
produce monopoles, then this divergence is zero.  Also note that
the field \b{$<\vec{C}>$} is generally assumed to be independent of
the field \b{$<\vec{E}>$}, which is a valid assumption for
biological media in general, except if ephaptic interactions must
be taken into account.

We also have
\b{
\begin{equation}
 <\vec{\j}>(\vec{r},t) = \int_{-\infty}^{+\infty}<\sigma^e>(\vec{r},\tau,\vec{E},\vec{H})~<\vec{E}>(\vec{r},t-\tau)~d\tau ~+ <D>~\nabla<\rho^{free}>.
\label{cap1}
\end{equation}}
 where \b{$\sigma^{e}$} is the electric conductivity and \b{$<D>$}
is the mean ionic diffusion tensor \footnote{{Because the law
of ionic diffusion is given by \b{$\vec{\j}_{mat}=-D\nabla C$} when
the units of \b{$C$} are \b{$mol/m^3$} and when we have only one
type of ion, we have multiplied the expression of
\b{$\vec{\j}_{mat}$} by \b{$-zF$} to yield the electric current
density \b{$\vec{\j}=D\nabla \rho $} associated to each ionic
species.  \b{z} is the valence of the ions considered, and
\b{$F=9.65\times 10^4~C/mol$} is the Faraday constant.  The choice
of the sign is according to the standard convention.  Note that if
the fundamental charge is taken as that of the proton, then one
must multiply by the factor \b{$zF$}, but if it is that of the
electron, then the multiplying factor is \b{$-zF$}.}}.  We define
\b{$<D>$} as follows:
\b{
$$
<D>{\sum_{i=1}^{N}\nabla <\rho_i^{free}>}=\sum_{i=1}^{N}<D_i>\nabla <\rho_i^{free}>
$$
}
where the sums run over the different ionic species.  Thus, we can write that the part of current
density caused by concentration changes equals \b{$<D>\nabla
<\rho^{free} >$}\footnote{Note that the spatial average of
\b{$<D>$} will have similar values for different ionic species
because the diffusion coefficients of the main ions
(\b{$k^+,~Na^+,~Cl^-,~Ca^{++}$)} have similar values for biological
tissues in physiological conditions (see \cite{Nicholson1998})}. In
expression (\ref{cap1}), we have separated the current produced by
ionic diffusion from the current produced by other physical causes
such as Ohm's effect, polarization, etc.  Note that this separation
was made here for simplicity, but it is also possible to integrate
diffusion effects in the expression of the mean conductivity (see
Eq.~\ref{rel1} in Section~\ref{activesec}).  

It is important to note that the first term in the righthand side
of Eq.~\ref{cap1} is not exclusively due to Ohm's law (which
relates to energy dissipation), but can reduce to it in some cases
\cite{BedDes2009a,Bed2006b}.  In general, Eq.~\ref{cap1} gives a
time-dependent electric conductivity (or frequency-dependent in
frequency space), which is not the case for Ohm's law in general
(see Appendix~\ref{appenB}).  Also note that the integrals in
Eqs.~\ref{lien1} and \ref{cap1} can be seen as convolution products
relative to time, in which case they take the form of a simple
product in frequency space.

Thus, according to this theory, it is sufficient to model the
physical and geometrical nature of the extracellular medium by
using electromagnetic parameters and diffusion coefficients to
simulate the interaction fields when the current sources are known
-- this is usually called the {\it forward problem}.  Inversely, we
can also deduce the physical characteristics of the sources
from the knowledge of the electromagnetic parameters and 
interaction fields, as well as their spatial and temporal 
variations -- this is known as the {\it inverse problem}.  

Finally, the integrals in Eqs.~\ref{lien1} and \ref{cap1} must
satisfy the causality principle, according to which the future
cannot determine the present state of the system.  For example, the
value of electric field \b{$<\vec{E}>$} at time \b{$t +|\Delta t|$}
must not influence the value of electric displacement
\b{$<\vec{D}>$} at time \b{t}.  Thus, the causality principle
determines a supplementary constraint on the possible values of
tensors \b{$<\mu>$}, \b{$<\varepsilon>$} and \b{$<\sigma^e>$},
which limits the number of possible mathematical models of the
extracellular medium. For instance, as detailed below in
Section~\ref{KK}, this principle imposes mathematical relations
between the electric parameters \b{ $<\varepsilon>$} and
\b{$<\sigma^e>$}, which are called Kramers-Kronig relations for
linear media.

The set of equations above define a mean-field formalism in which 
Maxwell equations are formulated with spatial averages.  In the next
sections, we consider different approximations to this formalism.

\subsection{The quasi-static approximation in mean field}

The first approximation to the Maxwell equations is the {\it
quasi-static approximation}, which consists of de-coupling electric
and magnetic variables.  In general, the time variation of
\b{$<\vec{B}>$} produces an electric field \b{$<\vec{E}>$}
(Lenz-Faraday effect), the electric and magnetic variables are
coupled in Maxwell equations (Eq.~\ref{Mwell}) by the following
expression: 
\b{
\begin{equation}
 \nabla\times<\vec{E}>~=~-\frac{\partial<\vec{B}>}{\partial t}
\end{equation}}

It was shown that for biological media and current sources, the
Lenz-Faraday effect is negligible \cite{Bosetti2008}.  In such
conditions, we can write: 
\b{
\begin{equation}
 \nabla\times<\vec{E}>~=~0
\end{equation}}

Under this quasi-static approximation, the electric and magnetic
variables are de-coupled in Maxwell equations, and the electric
field obeys:
\b{
$$
<\vec{E}> ~=~ -\nabla <V>
$$}

This approximation is also called the \textit{a priori} quasistatic
approximation, by opposition to the \textit{a posteriori}
quasistatic approximation, which consists of finding the general
solution of Maxwell equations and later de-couple the electric and
magnetic variables (see details in \cite{Bosetti2008}).  Although
these two approximations are not strictly equivalent, we will only
consider the \textit{a priori} quasistatic approximation in the
remaining of this paper.  

According to this approximation, Maxwell equations simplify to the
following expressions:
\b{
\begin{equation}
\left\{
 \begin{array}{rclccc}
  \nabla\cdot<\vec{D}> &=& <\rho^{free}> 
 \\\\
  \nabla\times<\vec{E}> &=& 0 \\\\
\nabla\cdot<\vec{\j}> +\frac{\partial<\rho^{free}>}{\partial t}&=&0
 \end{array}
\right .
\label{eqgenerale}
\end{equation}}
where the current density \b{$<\vec{\j}>$} is linked to the
electric field \b{$<\vec{E}>$} by Eq.~\ref{cap1}.   Note that, contrary to the static cases
(electrostatics and magnetostatics), the fields \b{$<\vec{E}>$},
\b{$<\vec{D}>$}, \b{$<\vec{\j}>$} and \b{$<\rho^{free}>$} are here space 
and time-dependent.  

In the following, we consider the complex Fourier transform
\b{
$$
X_{\omega} =\int_{-\infty}^{+\infty} X(t) \ e^{-i\omega t} dt~~~,~~~
X(t) =\frac{1}{2\pi}\int_{-\infty}^{+\infty} X_{\omega} \ e^{i\omega t} d\omega
$$}
where \b{$\omega=2\pi f$}.  Note that because of the linearity of 
the spatial average, we have 
\b{$<\vec{X}>_{\omega}=<\vec{X_{\omega}}>$}.

Applying the complex Fourier transform to Eqs.\ref{lien1}, \ref{cap1} and
\ref{eqgenerale} leads to:
\b{
\begin{equation}
\left\{
 \begin{array}{rcl}
  \nabla\cdot (<\varepsilon_{\omega}>\nabla< V_{\omega}>) &=& -<\rho_{e\omega}^{free}> +~\nabla\cdot<\vec{C}_{\omega}>
 \\\\
\nabla\cdot(<\sigma_{\omega}^{~e}>\nabla <V_{\omega}>)  &=&i\omega<\rho_{e\omega}^{free}>~-i\omega~\nabla\cdot<\vec{C}_{\omega}> +~\nabla\cdot (<D>~\nabla<\rho_{\omega}^{free}>)
 \end{array}
\right .
\label{freq1}
\end{equation}}

The first of these equations is Poisson's law in mean-field. 
Although in some cases (electrostatics), the Poisson equation is
sufficient to determine the solution of the system, it is not
sufficient in the quasistatic case, and the second equation is
necessary to close the system.  This second equation is the
differential law of charge conservation (in the presence of
diffusion) and takes into account the time variations of the
electric potential

By multiplying the first equation of Eqs.~\ref{freq1} by
\b{$i\omega$}, and adding the result to the second equation, leads
to:
\b{
\begin{equation}
\nabla^2<V_{\omega}> + \frac{\nabla<\gamma_{\omega}> }{<\gamma_{\omega}>}\cdot\nabla <V_{\omega}>
 = \frac{1}{<\gamma_{\omega}>}~\nabla\cdot (<D>\nabla<\rho_{\omega}^{free}>)
\label{CCC}
\end{equation}}
where \b{$\gamma_{\omega} =
<\sigma_{\omega}^{~e}>+i\omega<\varepsilon_{\omega}> $} is the
admittance of the extracellular medium.  This equation is general
and can be used to calculate the extracellular potential in an
extracellular medium with arbitrarily complex properties (i.e., 
when the electric parameters depend on frequency and space).  It 
is a generalization of expressions obtained previously
\cite{Bed2004,BedDes2009a}. The righthand term accounts for
ionic diffusion.

{Because Maxwell equations are scale invariant, the expression
above (Eq.~\ref{CCC}) is valid at all scales.  Like in any
mean-field approach, the spatial scale can be chosen according to
the scale of the phenomenon that needs to be modeled, as well as
the physical size and distance between electrodes. For example, in
the case of CSD of mammalian cerebral cortex, one must consider
scales of the order of \b{50~$\mu$m} to resolve the field produced
by each cortical layer.}

Note that in the quasistatic approximation, the explicit dependence
of the electric field on magnetic permeability \b{$\mu$} completely
disappears.  However, there can still be an implicit dependence
through \b{$\vec{H}$} in nonlinear media, because the electric
field does not depend explicitly on magnetic induction anymore. 

\subsection{The quasistatic approximation at larger scales}

At small scales ($\approx 1~\mu$m), biological media such as the
cerebral cortex are far from homogeneous and isotropic.  The
electric parameters can display large variations, for instance
between fluids and membranes.  However, at larger scales ($\approx
50~\mu$m), such media can be considered as homogeneous and
isotropic.  In such a case, the tensors \b{
$<\varepsilon_{\omega}>$} and \b{$<\sigma_{\omega}^{~e}>$} can
reduce to scalar quantities.  Note that the fact of considering
larger scales suppresses the directional dependence of the
propagation of currents by a statistical equivalent, without
changing the frequency dependence produced by physical phenomena at
small scales.  The transition from small scales to larger scales
gives the same form as Eqs.~\ref{CCC}, but with scalar parameters
which will have an explicit dependence on space, frequency and the
values of the field in general\footnote{Note that the space
dependence is much smaller than the large variations seen at
microscopic scales, for example between fluids and membranes.}. The
rate of spatial variation of these parameters at scales of the
order of \b{$50~ \mu$}m is approximately zero, such that:
\b{
\begin{equation}
\left \{
\begin{array}{ccc}
\nabla <\sigma_{\omega}^{~e} > |_{10^6~\mu m^3} & \approx & 0 \\\\
\nabla <\varepsilon_{\omega} > |_{10^6~\mu m^3} & \approx & 0
\end{array}
\right .
\end{equation}}
Note that there can be a frequency dependence of the current
propagation which results from microscopic inhomogeneities of the
electric parameters \cite{Bed2004}, from polarization phenomena
\cite{Bed2006b} or from ionic diffusion \cite{BedDes2009a}.  This
frequency dependence of the current will not disappear when
considering larger scales.  On the other hand, new frequency
dependencies may appear, such as for example the transformation of
a frequency-independent conductivity tensor at small scales
(\b{$\sim 1~\mu$m}) to a scalar conductivity at large scales
(\b{$\sim 50~\mu$m}) will be associated to a frequency dependence
of this macroscopic conductivity (for details, see
\cite{BedDes2009a}).  In agreement with this, measurements of the
macroscopic conductivity demonstrated strong frequency dependence
in different biological tissues \cite{Gabriel1996}.

\subsection{The linear approximation in mean-field}

Still within the quasistatic approximation, we now consider the
further simplification that the extracellular medium is {\it
linear}.  In a linear medium, the electric parameters are
independent of the values of the fields (note that this linearity is
different than that of Maxwell equations, which are always linear).
In this case, the electric parameters only depend on space and time
(or space and frequency).

In this case, the system of equations \b{(\ref{freq1})} becomes
linear at small scales.  This linear approximation is easy to
justify for the magnetic field, given the small currents involved
(for example, \b{$4\pi\times 10^{-7}~H/m$} in neocortex) and the
gradient of \b{$\mu$} is almost zero (see \cite{Hamalainen93}).  In
contrast, the linear approximation is less trivial in the case of
the electric field because of the many nonlinearities involved. 
For example, several ionic conductances are strongly
voltage-dependent (such as the Na$^+$/K$^+$ conductances involved
in action potentials), which will make the electric parameters of
membranes strongly dependent on the electric field.  Nevertheless,
the total volume of tissue occupied by membranes is small compared
to other regions where the linear approximation is valid, so
biological tissues can in general be considered as linear.  Note
that this linearity is evident for low frequencies ($<$10~Hz), but
it is less evident for high frequencies ($>$100~Hz), where
nonlinear phenomena such as action potentials can have a major
contribution.

\subsection{The Kramers-Kronig relations under the linear
approximation} \label{KK}

As discussed above in Section~\ref{MF}, the causality principle
determines a supplementary constraint on the possible values of
tensors \b{$<\mu>$}, \b{$<\varepsilon>$} and \b{$<\sigma^e>$}.  In
the linear approximation, one can show that, in general (for
isotropic media of class \b{$C_2$}), the linking equation between
the electric displacement and electric field takes the following
form:
\b{
\begin{equation}
 \vec{D}(\vec{x},t) = \vec{E}(\vec{x},t) + \int_0^{\infty}f(\vec{x},\tau)\vec{E}(\vec{x},t-\tau)~d\tau
\end{equation}}

In this case, one can show that the frequency dependence of
electric parameters is not arbitrary but is linked by the
Kramers-Kronig relations (see Section 82 in \cite{Landau1981}):
\b{
\begin{equation}
\begin{array}{ccccc}
\varepsilon_{\omega}(\vec{x})-\varepsilon_{\infty}(\vec{x}) &=&  
\frac{2}{\pi} \fint_{0}^{\infty}\frac{~\sigma_{\omega'}^{~e}(\vec{x})-\sigma_0^{~e}(\vec{x}) }{\omega^{'2}-\omega^2}~d\omega'\\\\
\sigma_{\omega}^{~e}(\vec{x}) -\sigma_0^{~e}(\vec{x}) &=& - \frac{2\omega^2}{\pi} \fint_{0}^{\infty}
\frac{\varepsilon_{\omega'}(\vec{x})-
\varepsilon_{\infty}(\vec{x})}{\omega^{'2}-\omega^2}~d\omega'
\label{kronig}
\end{array}
\end{equation}}
where principal integrals (\b{$\fint$}) are used. 
\b{$\varepsilon_{\infty}$} is the absolute electric permittivity of
vacuum and \b{$\sigma_0^{~e}$} is the static electric conductivity
(\b{$\omega=0$}).  Note that these relations can be seen as a
direct and inverse transform.  The Maxwell-Wagner time
\b{$|\frac{\varepsilon_{\omega}}{\sigma_{\omega}^{~e}}|$}
represents the characteristic time (or ``inertia'') for settling
into a stationary regime, and can be strongly frequency dependent
\cite{Bed2006b}.  Interestingly, this ratio is mathematically
analogous to the time-frequency uncertainty principle in Fourier
transforms when the electric conductivity of the extracellular
medium becomes very small at zero frequency (see
Appendix~\ref{kk}).  Note that there exists no such relation for
the spatial variations of electric parameters, which are specific
to each medium.

It is important to note that the Kramers-Kronig relations have a
strong consequence on the plausibility of purely resistive media. 
If a medium is purely resistive, then both conductivity and
permittivity are constant and independent of frequency.  However,
if one takes into account a very weak frequency dependence of
conductivity (``quasi-resistive'' media), such as
\b{$\sigma_{\omega}^{~e} -\sigma_0^{~e} \sim f^\alpha$} with
\b{$\alpha<<1$}, then the Kramers-Kronig relations impose that we
necessarily have \b{$\varepsilon_{\omega}-\varepsilon_{\infty} 
\sim f^{-(1-\alpha)}$}.  In such a case, the permittivity will be
strongly frequency dependent, so will be the Maxwell-Wagner time
\b{$\tau_{\mbox{\tiny
MW}}=|\frac{\varepsilon_{\omega}}{\sigma_{\omega}}|$}.  Thus, a
purely resistive extracellular medium is a singularity, and is not
likely to be realistic model for complex biological media.

\section{Application of the quasistatic mean-field theory
to linear media}

In this section, we apply the theory outlined above to media which
are linear in the electromagnetic sense, and which are homogeneous
and isotropic (at macroscopic scales).  This is equivalent to
assume that the macroscopic parameters \b{$<\sigma_{\omega}>$},
\b{$<\varepsilon_{\omega}>$} and \b{$<D>$} are scalars independent
of space, such that
\b{$\nabla(<\sigma_{\omega}>+i\omega<\varepsilon_{\omega>}) \approx
0$} and \b{$\nabla <D>\approx 0$}.  This approximation is certainly
valid for relatively large distances (greater than $\sim 50~\mu
m$).  We also consider the system under the quasistatic
approximation as defined above.

In the following sections, we examine different limit cases.  The
first case corresponds to the standard model with dipolar sources
and a resistive (or quasi-resistive) extracellular medium.  A
second case will consider the same model, but with additional
monopolar sources.  The third case will consider ionic diffusion
(not present in the two first cases), which also implies monopolar
sources.  In each case, we will derive the expression to use for
CSD analysis.

\subsection{Dipole sources in resistive and quasi-resistive media}
\label{stand}

We start with the standard model in which the electric conductivity
of the extracellular medium is constant (in space and frequency),
scale invariant, and isotropic. We also consider that diffusion is
negligible.  Under these hypotheses, we have \b{$\nabla
<\gamma_{\omega}>=0$} (homogeneous and isotropic medium) and
\b{$<\gamma_{\omega}> = \bar{\sigma}=cst.$}, and the general
formalism (Eq~\ref{CCC}) reduces to:
\b{
\begin{equation}
 \nabla^2<V_{\omega}> 
 ~ = ~0
\label{mod1}
\end{equation}
}

We can also consider a slightly more realistic model of the
extracellular medium by assuming that it is quasi-resistive
(\b{$<\sigma_{\omega}>\approx\bar{\sigma}$}) instead of resistive
(\b{$<\sigma_{\omega}>=\bar{\sigma}$}) (because the latter
represents a singularity as outlined above in Section~\ref{KK}),
isotropic and homogeneous for large scales (\b{$\sim 50~\mu$m};
the medium is allowed to be non-homogeneous for smaller scales of
the order of \b{$\sim 1~\mu$m}).  Note that in general, the
non-homogeneity of conductivity at smaller scales can induce a
frequency dependence at larger scales (see
\cite{Bed2004,BedDes2009a}).  Thus, the hypothesis that the
macroscopic conductivity is independent of frequency is equivalent
to assume that there is no significant variations of impedance at
microscopic scales.  

In such conditions, Eq.~\ref{CCC} again reduces to:
\b{
\begin{equation}
 \nabla^2<V_{\omega}>~ 
 =~ 0
\label{mod}
\end{equation}
}

{In some formulations, the standard model} does not consider 
the possibility of microscopic ($\sim 1~\mu$m) monopolar
sources~\cite{Mitzdorf,Pettersen}.  This is equivalent to
hypothesize that, at every time, each portion of cell membrane has
an equal number of positive and negative charges at opposite sides
of the membrane, such that it is locally neutral.  This hypothesis
is also equivalent to state that the extracellular electric field
is produced by dipoles (or more complex multipolar arrangements),
and that the monopolar component of the field is negligible at
scales of $\sim 1~\mu$m\footnote{Note that one cannot say that the
monopolar component is rigorously zero, because there is at least a
monopolar component in the ion channels themselves, because of ion
selectivity.}.  This implies that the attenuation of the
extracellular potential follows a law which varies as \b{$1/r^2$}
(or \b{$1/r^3$}, \b{$1/r^4$} ... for multipoles of higher order)
when \b{$r\rightarrow\infty$}, where \b{$r$} is the distance to the
source.  Thus, in the standard model, the electric displacement in
frequency space (see Eqs.~\ref{lien1} and \ref{deplac}) is given by:
\b{
\begin{equation}
 <\vec{D}_{\omega}^*> ~=~ \varepsilon_{\infty}<\vec{E}_{\omega}>
+<\vec{P}_{\omega}> + <\vec{C}_{\omega}>
~=~<\varepsilon_{\omega}><\vec{E}_{\omega}>+ <\vec{C}_{\omega}>
\end{equation}}
where
\b{
\begin{equation}
\left \{
\begin{array}{rcl}
\nabla\cdot<\vec{D}_{\omega}^*>&=&0\\\\
\nabla\cdot\varepsilon_{\infty}<\vec{E}_{\omega}>&=&
+<\rho_{\omega}^{bound}>\\\\
 \nabla\cdot<\vec{P}_{\omega}>&=&-<\rho_{\omega}^{bound}>\\\\ 
\nabla\cdot<\vec{C}_{\omega}>&=&
0
\end{array}
\right .
\end{equation}}
at large scales (\b{$\sim 50~\mu$m}). Taking the inverse
Fourier transform, one obtains:
\b{
\begin{equation}
\nabla^2<V> ~ =~ 0
\label{csd2}
\end{equation}} 

According to this model, the inverse solution (CSD method) can be
obtained assuming that the voltages measured at $n$ different
extracellular sites are solution of Laplace equation.  According to
the superposition principle, the extracellular potential can be
considered as resulting from a sum of $n$ macroscopic dipolar
sources for sufficiently large $n$.  Note that the value of $n$ is
determined by Shannon's sampling theorem, according to which the
number of samples (number of electrodes $n$) must be twice larger
than the larger spatial frequency of the field.  To evaluate these
$n$ dipolar sources, one can simply apply the inversion of the
matrix linking the $n$ measured voltages with the $n$ dipolar
sources according to the ``forward'' solution of Laplace equation
(see e.g., \cite{Pettersen}).  Note that this approach is different
than the classic CSD method proposed by Mitzdorf~\cite{Mitzdorf},
which is based on a Poisson type equation.

The hypothesis of local neutrality in a homogeneous and isotropic
extracellular medium implies that the frequency dependence of the
measured signal is only due to the frequency dependence of the
source (for example the effect of morphology -- see
\cite{Linden2010}, the exponential or bi-exponential nature of
synaptic conductances, correlations in synaptic activity, action
potentials, etc), because Laplace equation does not explicitly
depend on frequency.  Thus, according to the standard model, there
is no filtering due to extracellular space and the power spectrum
of the extracellular potential is identical to that of the current
sources.

Finally, it is important to note that, in a resistive extracellular
medium, if we express the extracellular potential as a function of
the dipole moments instead of the current sources, then the power
spectral density (PSD) of the electric potential will necessarily
have a supplementary frequency dependence of the form
\b{$1/\omega^2$} compared to that of the current. This is due to
the fact that the current is proportional to the temporal
derivative of the the dipole moment \b{$\vec{p}_{\omega}$} (defined
from the charge distribution). However, the situation is different
if the medium is quasi-resistive.  In this case, the Kramers-Kronig
relations give \b{$\varepsilon_{\omega}\sim\frac{1}{\omega}$}, and
thus the ratio \b{$\frac{\vec{p}_{\omega}}{\varepsilon_{\omega}}$}
will have little frequency dependence.  It follows that the PSD of
the extracellular potential will have approximately the same
frequency dependence as the current sources in a quasi-resistive
medium.  This is a striking difference between resistive and
quasi-resistive media.  As discussed above, the latter is a more
realistic situation because any spatial variation of microscopic
conductivity will necessarily induce a frequency dependence of the
macroscopic conductivity (see \cite{Bed2004}).

\subsection{Monopolar sources in resistive and quasiresistive media}
\label{monop}

In the previous section, we hypothesized that the extracellular
medium is locally neutral at microscopic scales ($\sim 1~\mu$m),
and thus, that the sources of the electric potential are dipoles.
We now relax this hypothesis, and allow significant electric
monopoles to appear in addition to conductance variations, so that
the field results from both monopolar and dipolar contributions. 
Electric monopoles could result from different physical sources,
such as the ionic selectivity of synaptic ion channels (similar to
a ``Maxwell Daemon''), combined with the finite velocity of charge
movement~\cite{BedDes2008}.  These factors should create some
accumulation of charge in the vicinity of the synapse when synaptic
conductances are activated\footnote{Note that it is important here
to take into account the {\it spatial extent} of the synapse,
contrary to the standard theory where synapses are considered as
point processes.}.  Note that monopoles are transient by
definition, and equivalently, one could consider that the
conductance variations determine a non-stationary regime
\b{$\nabla\cdot<\vec{\j}> + \frac{\partial ~<\rho^{free}>}{\partial
t}=0$} (see Appendix~\ref{CC}).  In this transient regime,
Kirchhoff's ``point rule'' does not apply (it is based on the law
of current conservation \b{$\nabla\cdot\vec{\j}=0$}) and would
apply only when the system reaches a stationary state.  However,
Kirchhoff's ``loop rule'' is always valid under the quasistatic
approximation, because we have \b{$\nabla\times\vec{E}=0$}, and
consequently \b{$\oint\vec{E}\cdot\vec{ds}=0$}, which is at the
basis of the latter rule.

{Contrary to the assumptions of the dipole model, monopolar
sources imply that integrating the charge density over a closed
surface surrounding each source is non-zero.  To include the
contribution of monopolar current sources, we have} 
\b{
\begin{equation}
 I_{\omega}^n = \oiint\limits_{\partial D} <\vec{\j}_{\omega}>\cdot\hat{n}~dS
=\iiint\limits_{D}\nabla\cdot<\vec{\j}_{\omega}>dv=
-i\omega Q_{\omega}\neq 0
\label{courant}
\end{equation}}
where \b{$Q_{\omega}$} the total charge contained in the source. 
Note that this relation shows that the monopolar component is
linked to the current through a temporal derivative, which is a
consequence of the charge conservation law.  Consequently, the
extracellular potential (which is here proportional to the charge)
will not have the same power spectrum as the source, and will have
an additional \b{$\sim \frac{1}{\omega^2}$} component for a
resistive extracellular medium.  However, similarly to the case of
dipolar sources in previous section, the situation is different for
a quasi-resistive medium.  The Kramers-Kronig relations imply
\b{$\varepsilon_{\omega}\sim\frac{1}{\omega}$}, and the ratio
\b{$\frac{Q_{\omega}}{\varepsilon_{\omega}}$} will have very little
frequency dependence and the PSD of the extracellular potential
will be very similar to that of the sources.

If we consider the same conditions as for the standard model
(resistive or quasi-resistive media), we obtain
\b{
\begin{equation}
\left \{
\begin{array}{rcl}
\nabla\cdot<\vec{D}_{\omega}^*>&=& <\rho_{e\omega}^{free}>\\\\
\nabla\cdot\varepsilon_{\infty}<\vec{E}_{\omega}>&=&
<\rho_{e\omega}^{free}>+<\rho_{\omega}^{\Delta~cond}>+<\rho_{\omega}^{bound}>\\\\
 \nabla\cdot<\vec{P}_{\omega}>&=&-<\rho_{\omega}^{bound}>\\\\ 
\nabla\cdot<\vec{C}_{\omega}>&=&- <\rho_{\omega}^{\Delta~cond}>
\end{array}
\right .
\end{equation}}

In such conditions, Eq.~\ref{CCC} becomes:
\b{
\begin{equation}
\nabla^2<V_{\omega}>~ =~0
\label{resistif}
\end{equation}}
for resistive and quasi-resistive extracellular media.  Thus, in 
temporal space, we have the same equation for both cases:
\b{
\begin{equation}
\nabla^2<V>~ =~0
\end{equation}}

{However, if we take into account monopolar current sources
and the law of charge conservation, then we have in general:}
\b{
\begin{equation}
\sigma_{\omega}\nabla^2<V_{\omega}>~ =~i\omega<\rho_{\omega}>
\label{resistif2}
\end{equation}}
where \b{$\sigma_{\omega} = cst$.}

Thus, the model with monopolar current sources has {a
different structure than the dipole model in Section~\ref{stand}
because \b{$<\rho_{\omega}>\neq 0$}}. Local neutrality in a
homogeneous and isotropic extracellular medium implies an identical
frequency dependence of the {current source \b{$I_{\omega} = -
i\omega <\rho_{\omega}>$})} and the extracellular potential.  Like
the standard model, there is no ``filter'' in this case.  There is
a notable difference, however.  The law of attenuation with
distance varies here in \b{$1/r$} instead of \b{$1/r^2$} for \b{$r
\to \infty$}.  If the number of electrodes is large enough to
respect Shannon's sampling theorem, then the current source
densities can be simply evaluated by approximating the Laplace
equation using finite difference methods, as well as the knowledge
of the ``forward'' solutions of this equation (see
\cite{Pettersen}). We will see in the next section that these
conclusions are different if ionic diffusion is taken into account.

\subsection{Models with ionic diffusion} \label{diffusion}

While the influence of ionic diffusion was neglected in the
previous sections, we now consider this case more explicitly
without any other hypothesis about the medium.  If a selective ion
channel opens, the flow of ions may induce accumulation of charges
in the region adjacent to the channel if ions diffuse faster than
the time needed for passing through the channel (which will
generally be the case).  The electric field resulting from
conductance variations is not selective on the type of ion, such
that the positive ions are attracted and negative ions are repulsed
if the field is negative (and {\it vice-versa} for a positive
field).  This is contrary to the flow inside the channel because it
is selective to only a subset of ionic species.  The combination of
these factors makes it unavoidable that there will be charge
accumulation around open ion channels.  In the standard model, this
charge accumulation is considered as negligible. 

We now evaluate the consequences of this phenomenon on the
frequency dependence of the field produced by ionic conductances in
the subthreshold regime.  If we consider a homogeneous
extracellular medium with constant electric parameters (independent
of frequency at large scales, $\sim 50~\mu$m), then we have:
\b{
\begin{equation}
\begin{array}{rcl}
<\sigma_{\omega}^{~e}>|_M &=& \bar{\sigma}\\\\
<\varepsilon_{\omega}>|_M  &=& \bar{\varepsilon}\\\\ 
<D>|_M &=& \bar{D} \neq 0
\end{array}
\end{equation}}
where the parameters \b{$\bar{\sigma}$}, \b{$\bar{\varepsilon}$}
and \b{$\bar{D}$} do not depend on space.

According to those hypotheses, variations of ionic concentrations
appear in the vicinity of the open ion channels, and these
variations are opposite to the current produced by the electric
field resulting from conductance variations. It thus appears that
the conditions of current propagation at microscopic scales ($\sim
1~\mu$m) cannot fulfill the condition of homogeneous ion
concentration which is at the basis of Ohm's differential law (see
Appendix~\ref{appenB}).  In such conditions, the electric
parameters of the extracellular medium have the following form at
microscopic scales ($\sim 1~\mu$m):
\b{
\begin{equation}
\begin{array}{rcl}
<\sigma_{\omega}^{~e}>|_m(\vec{x}) &=& \bar{\sigma}_m(\vec{x})\\\\
<\varepsilon_{\omega}>|_m (\vec{x}) &=& \bar{\varepsilon}_m(\vec{x})\\\\ 
<D>~|_m(\vec{x}) &=& \bar{D}_m \neq 0
\end{array}
\end{equation}}
with 
\b{
\begin{equation}
 <\gamma_{\omega}>|_M \ = \ <\mbox{\begin{small}$<\gamma_{\omega}>|_m$
 \end{small}}>|_M
\label{cond}
\end{equation}
}
This last equation is necessary to keep the consistency between
microscopic ($\sim 1~\mu$m) and macroscopic ($\sim 50~\mu$m)
scales. 

According to this model, the current density (at microscopic
scales, $\sim 1~\mu$m) is given by:
\b{
\begin{equation}
 <\vec{j}>|_m = -\bar{\sigma}_m\nabla <V>|_m +\bar{D}_m~\nabla <\rho>|_m
\label{cour1}
\end{equation}
}
where we have (see Eq.~\ref{ohm1})
\b{
\begin{equation}
\bar{\sigma}_m = \lambda_q\tau_cn_v(\vec{x},t)
\label{sig}
\end{equation}
}
This expression can be deduced by separating the domain into 
sufficiently small elements such that ion density can be considered
as spatially homogeneous, and sufficiently large for Ohm's law to 
apply.

According to Boltzmann distribution (see Appendix~\ref{activesec}),
we have
\b{
\begin{equation}
\nabla<\rho>|_m(\vec{x},t)=\frac{<q>^2}{kT}~n_v\nabla <V>|_m(\vec{x},t)=\frac{<q>^2}{k\lambda_q\tau_cT}\bar{\sigma}_m\nabla <V>|_m
\end{equation}}
By taking into account Eqs.~\ref{cour1} and \ref{sig}, we obtain
\b{
\begin{equation}
 <\vec{j}>|_m = [ \bar{D}_m-\frac{k\lambda_q\tau_cT}{<q>^2}]~\nabla <\rho>|_m
=<\beta>|_m~\nabla<\rho>|_m
\label{42}
\end{equation}}
where \b{$<\beta>$} is an effective diffusion coefficient.  Note
that the value of \b{$<\beta>$} is smaller than the mean diffusion
coefficient because \b{$\frac{k\lambda_q\tau_cT}{<q>^2} $} must be
positive.  The value of \b{$<\beta>$} also depends on the values of
ionic concentrations because several parameters in Eq.~\ref{42} are
concentration-dependent and is proportional to temperature because
the ionic diffusion coefficient is itself proportional to
temperature (see for example the Einstein relation for diffusion).

Applying the differential law of charge conservation, we get
\b{
\begin{equation}
 <\beta>|_m ~\nabla^2 <\rho>|_m =- \frac{\partial<\rho>|_m}{\partial t}
\end{equation}}
Thus, the charge density produced in the vicinity of the ion
channel is solution of a parabolic differential
equation similar to the diffusion equation\footnote{
% Note that
% contrary to a classic diffusion equation, the coefficient
% \b{$\beta$} is not necessarily positive, because the parameter
% \b{$kT\frac{\lambda_q\tau_c}{q}\approx kT\frac{\tau_c}{<m>}$}
% varies between \b{$ 10^{-7}-10^{-10} ~m^2/s$} for \b{$<m>\approx
% 10^{-25}-10^{-26}~kg$}, \b{$\tau_c\approx 10^{-12}-10^{-14}~s$} and
% \b{$T=300~^oK$}, with \b{$\bar{D}_m\approx 10^{-9}~ m^2/s$}.  Also
Note that the coefficient \b{$<\beta>|_m$} depends on ion concentrations
via \b{$\lambda_q$}, \b{$<q>^2$} and \b{$\tau_c$}, and thus could
vary greatly according to the activity of the surrounding neurons.}.

It follows that the charge density obeys: 
\b{
\begin{equation}
 \nabla^2 <\rho_{\omega}>|_m  =-i\frac{\omega}{<\beta>|_m}<\rho_{\omega}>|_m
 \label{chardens}
\end{equation}}

At microscopic scales ($\sim 1~\mu$m), we obtain 
{(see Eq.~\ref{CCC})}: 
\b{
\begin{equation}
\nabla^2<V_{\omega}>|_m + \frac{\nabla(<\gamma_{\omega}>|_m) }{<\gamma_{\omega}>|_m}\cdot\nabla <V_{\omega}>|_m =\frac{i\omega}{<\gamma_{\omega}>|_m}\cdot
\frac{<D>|_m}{<\beta>|_m}
 <\rho_{\omega}>|_m \sim i\omega<V>|_m
\label{microscopique}
\end{equation}}
Here, the proportionality between \b{$<\rho_{\omega}>|_m$} and
\b{$<V_{\omega}>|_m$} can be deduced from the linear (first-order)
approximation of Eq.~\ref{Bolz1a} (see Appendix~\ref{activesec}). 
The second-order approximation would give a cubic term in
\b{$<V_{\omega}>|_m$}.  

Applying the consistency equation between scales by assuming the
statistical independence of the parameters leads to the following
equality:
\b{
\begin{equation}
\nabla^2<V_{\omega}>|_M =-\frac{i\omega}{<\gamma_{\omega}>|_M}\cdot
\frac{1}{<\beta>|_M}
 <\rho_{\omega}>|_M 
\label{foufond}
\end{equation}}
with
\b{
\begin{equation}
\begin{array}{rclcc}
\frac{1}{<\gamma_{\omega}>|_M} &=& \frac{1}{N}\sum\limits_{j=1}^N \frac{1}{<\gamma_{\omega}^j>|_m}   \\\\
\frac{1}{<\beta>|_M} &=&\frac{1}{N}\sum\limits_{j=1}^N
\frac{<D^j>|_m}{<\beta^j>|_m}\\\\
<D>|_M &=&\frac{1}{N}\sum\limits_{j=1}^N
<D^j>|_m
\end{array}
\end{equation}}
where $N$ is the ratio between the reference volumes at macroscopic
and microscopic scales (note that to simplify the formalism, we
have approximated the macroscopic mean by a discrete summation over
microscopic means). The second term of the lefthand side of
Eq.~\ref{microscopique} becomes zero at macroscopic scales (see
consistency equation Eq.~\ref{cond}).  Note that the means over
parameters \b{$\gamma_{\omega}$} and \b{$\beta_m$} are harmonic means,
while the means over matter fields are arithmetic means.

Finally, by applying the inverse Fourier transform, we obtain (for
Maxwell-Wagner times much smaller than unity): 
\b{
\begin{equation}
\bar{\sigma}~\nabla^2<V>|_M =
 -\frac{1}{<\beta>|_M}\frac{\partial}{\partial t}<\rho>|_M
\label{csd4}
\end{equation}}

Thus, the CSD method in the presence of ionic diffusion takes a
form which is very close to the Mitzdorf model \cite{Mitzdorf},
because we have one source term.  However, there are two notable
differences: first, the frequency dependence of charge density
implies that the extracellular medium will be frequency dependent
according to an impedance which varies as \b{$1/\sqrt(\omega)$}
{(see Appendix~\ref{Warburg})}.  Second, the extracellular
potential attenuates with distance according to a Yukawa potential
\b{$\frac{e^{-k(\omega)r}}{r}$} instead of \b{$\frac{1}{r^2}$}, as
in the standard model.  In this case, we have 
\b{
\begin{equation}
 |<V_{\omega}>|_m (r)| = |<V_{\omega}>|_m (R)~|
  \frac{Re^{-\frac{1}{2}\sqrt{\frac{\omega}{|<\beta>|_m|}}(r-R)}}{r}
\end{equation}}

Here, the extracellular potential is proportional to the charge
density (see Eqs.~\ref{Bolz1a}, \ref{jjj-1}, and \ref{jjj} in
{Appendix~\ref{Warburg})} under the linear approximation and for a
spherical source. It is interesting to note that the exponential
term increases with frequency such that the extracellular medium
favors the propagation of low frequencies (low-pass filter), as
shown in Fig.~\ref{appYukawa}).  This type of attenuation law in
Fourier space is also consistent with an exponentially-decaying
impedance.  If the frequency spectrum is narrow, one can replace
\b{$\omega$} by its maximal value
\b{$k=\frac{1}{2}\sqrt{\frac{max(\omega)}{|<\beta>|_m}|}$}, which
leads to an attenuation law for the potential as
\b{$\frac{e^{-k(r-r_o)}}{r}$}.  We can thus write that the electric
field is approximately equal to \b{$<\vec{E}>|_m=-\nabla
<V>|_m=e^{-k(r-r_o)}(kr+1)\frac{1}{r^2}\hat{r}$}.  In a resistive
medium, this leads to an electric resistivity given by
\b{$<\rho^e>_m=\frac{1}{<\sigma^e>|_m (r)} =\frac{4\pi}{I}e^{-k(r-r_o)}(kr+1) $}
(note that the current \b{$I=|\vec{j}|~4\pi r^2$} is conserved in a
resistive medium, and the field is given by \b{$<\vec{E}>|_m =
<\vec{j}>|_m / <\sigma(r)>|_m$}). Note that this particular
distance profile of the potential was calculated by assuming that
the medium is homogeneous {(see Appendix~\ref{Warburg})}, which makes
it applicable only at short distances from the membrane (of the
order of 10 to 50~nm).

\centerline{\note{[ -------------- Figure~\ref{appYukawa} here  -------------- ]}}

Finally, it is important to note that this frequency dependence
cannot be removed because it is an effect of the feedback caused by
ionic diffusion when ion channels open, and this is inherent to
biological tissue.  Because the PSD of the extracellular voltage is
of the form \b{$\sim \frac{1}{\omega} \ I(\omega)$}  {(see
Eq.~\ref{jjj} in Appendix~\ref{Warburg})}, one can view the effect
of ionic diffusion as a ``1/f filter'', as found previously
\cite{BedDes2009a}.

\subsection{Comparison between the different models}

We now compare the different cases examined here.  From the point
of view of the differential equations involved, in the ``standard''
model based on dipoles, {as well as with monopoles}, the
extracellular potential is solution of the Laplace equation, which
is elliptic.  In the third model with ionic diffusion, the
extracellular potential is solution of a Poisson type equation
where the source term is proportional to the time derivative of the
voltage (under the linear approximation), which gives a parabolic
equation.  As outlined above, the diffusion model is closer to the
monopole model (as diffusion can have monopolar effects) in a
resistive extracellular medium, but leads to a fundamentally
different mathematical form.  The physical reason for this
difference is that the ionic diffusion at the interface ion
channel/medium increases the inertia of the system as a function of
frequency.  
 
At the point of view of the CSD analysis method, different
algorithms must be used according to which model of the
extracellular medium is assumed.  In the two first cases, one must
use a ``forward'' solution because Laplace equation is non
invertible.  In this case, it is necessary to explicitly include
the distance dependence of the extracellular potential , which
varies as \b{$1/r^2$} for dipoles (for distances sufficiently large
compared to the size of the dipole) and \b{$1/r$} for the model
based on monopoles. Note that if the distance to the sources is not
large enough (compared to the typical size of the sources), or if
the dipolar moments are very large compared to monopolar moments,
then the attenuation will be closer to a linear combination of
\b{$1/r$} and \b{$1/r^2$}.  

In the diffusive model, however, the approach is totally different
because of the parabolic nature of the equations.  In this case, it
is enough to apply the Laplace operator to recover the sources. 
Two strategies are possible.  First, one could simply apply Laplace
operator on the extracellular voltage to yield estimates of the
current source densities.  Second, one could use a ``forward''
model and consider an attenuation law following a Yukawa potential
\b{$\frac{e^{-k(\omega)r}}{r}$} and apply the same procedure as for
the other models.

Perhaps the most interesting aspect is that the three different
models considered here have a different spectral signature.  In the
dipole model (or monopole model in a quasi-resistive medium), the
PSD of the extracellular potential is identical to that of the
sources.  The resistive monopole and the resistive dipole model
exert a filtering effect of \b{$1/f^2$} type, whereas the diffusive
model is equivalent to a \b{$1/f$} filter.  Thus, the frequency
characteristics of the signal can serve as a criterion to determine
the most appropriate model.  For example, if the PSD of the
extracellular voltage has \b{$1/f$} structure, this automatically
discards a pure monopole model, as well as dipole models in
resistive or quasiresistive media, and would suggest diffusive type
models.

\subsection{{Synthesis and applications to experimental data}}

{In this section, we synthesize the theoretical developments
provided here, and suggest a guide of how to apply them to
experimental data.  The generalization of the CSD method for
different cases of current sources and type of extracellular
medium, is summarized in Table~1.  The table considers monopolar
and dipolar current sources, as well as different types of
resistive and non-resistive media.}

{To perform a CSD analysis by allowing non-resistive
properties of the extracellular medium, we suggest the following
procedure.}

{\begin{enumerate} 
\item Estimate the type of extracellular medium from the power
spectral structure of CSD signals.  As detailed above
(Sections~\ref{stand}, \ref{monop}, \ref{diffusion}, the type of
medium (resistive, quasi-resistive, diffusive, etc) and type of
current sources (monopolar, dipolar, etc) can be inferred from the
power spectral structure of the extracellular potential.  This
analysis should be done on non-filtered data to set constraints on
the possible combinations of sources type of medium.
\item Identify the correct CSD expression compatible with the type
of source/medium inferred from power spectra.  Table~1  summarizes
the different cases considered here.  The expression identified is
then used to calculate the current sources from the extracellular
potential recordings.
\end{enumerate}}

\begin{table}[h!]
\centering
{
\begin{tabular}{|c|c|c|c|c|c|}
\hline
 & & & & & \\
Source  & Medium & $\varepsilon_{\omega}$ & $\sigma_{\omega}^e$ &  $\frac{V_{\omega}}{I_{\omega}^S}$ & Law \\
 & & & & & \\
\hline\hline
  %MONOPOLE
 & & & & & \\
1-pole  &  res.& $cst$& $cst$ & $\sim \frac{1}{r\omega}$& $\sigma_{\omega}^e\nabla^2<V_{\omega}>= i\omega<\rho_{\omega}^{free}>$
\\
\cline{2-6}
& & & & & \\
& quasi-res. & $\sim\frac{1}{\omega}$   & $cst$  & $\sim \frac{1}{r}$ &$\sigma_{\omega}^e\nabla^2<V_{\omega}>=i\omega<\rho_{\omega}^{free}>$
\\
\cline{2-6}
& & & & & \\
        & res.+dif.     & $\sim \frac{1}{\sqrt{\omega}}$   & $\sim \frac{1}{\sqrt{\omega}}$  & $\sim\frac{e^{-f(\omega)(r-r_o)}}{r\sqrt{\omega}}$ & $\nabla^2<V_{\omega}>= -i\omega\frac{<D>|}{<\gamma_{\omega}>|<\beta>|}<\rho_{\omega}^{free}>$ 
\\
\cline{2-6}
& & & & & \\
& gen.    & $\varepsilon_{\omega}$   & $\sigma_{\omega}^e$  & gen. & $\nabla^2<V_{\omega}>+\frac{\nabla<\gamma_{\omega}>}{<\gamma_{\omega}>}\cdot\nabla<V_{\omega}>= \frac{1}{<\gamma_{\omega}>}~\nabla\cdot (<D>\nabla<\rho_{\omega}^{free}>)$  \\
 & & & & & \\
\hline % DIPOLE
 & & & & & \\
\centering{2-pole}& res.      & $cst$ & $cst$& $\sim \frac{1}{r^2\omega}$ & $\nabla^2<V_{\omega}>= 0$ 
\\
\cline{2-6}
& & & & & \\
        & quasi-res.     & $\sim\frac{1}{\omega}$ & $cst$& $\sim \frac{1}{r^2}$ & $\nabla^2<V_{\omega}>= 0$
\\
\cline{2-6}
& & & & & \\
        & res.+dif.& $\sim\frac{1}{\sqrt{\omega}}$ & $\sim\frac{1}{\sqrt{\omega}}$ & $\sim\frac{e^{-f(\omega)(r-r_o)}}{r^2\sqrt{\omega}}$
& $\nabla^2<V_{\omega}>= 0$ 
\\
\cline{2-6}
& & & & & \\
& gen.     & $\varepsilon_{\omega}$   & $\sigma_{\omega}$  & gen. & $\nabla^2<V_{\omega}>+\frac{\nabla<\gamma_{\omega}>}{<\gamma_{\omega}>}\cdot\nabla<V_{\omega}>= 0$  
\\
 & & & & & \\
\hline\hline
& & & & & \\
\centering{Mitzdorf (2-pole)} &  res.     & $cst$ & $cst$& $\sim \frac{1}{r^2}$ & $\sigma^e\nabla^2<V> =-I_m$ 
\\
% & & & & & \\
% \centering{Mod\`ele de Pettersen} & r\'esistif     & $cst$ & $cst$& $\sim 1/r^2$ & $\sigma\nabla^2<V> =0$ \\
\hline
\end{tabular} }

\caption{{Different generalizations of the CSD method.  The
table shows the mean-field equations for different types of media,
and for monopolar or dipolar sources.  The Mitzdorf model is shown
apart, because it does not correspond to any of these mean-field
scenarios.  Abbreviations: res.  $\Rightarrow $ resistive
homogeneous medium, quasi-res.$\Rightarrow $ quasi-resistive
homogeneous medium, res. + dif.  $\Rightarrow $ resistive
homogeneous medium + ionic diffusion, gen.  $\Rightarrow $ general
(non-homogeneous, with spatial and frequency-dependent variations
of electric parameters), 1-pole $\Rightarrow $ monopole, 2-pole
$\Rightarrow $ dipole. Note that the frequency-dependence of the
permittivity and conductivity are not independent but are linked by
the Kramers-Kronig relations.  The quantity
\b{$\frac{V_{\omega}}{I_{\omega}^S}$} is the ratio between the
Fourier transform of the extracellular potential \b{$V_{\omega}$}
and the Fourier transform of each point current-source
\b{$I_{\omega}^S$} which produce the field (asymptotic solution,
far from the sources).  The function \b{$f(\omega)$} in ``res+dif''
determines a Yukawa type potential (see Fig.~\ref{appYukawa}).  }} 

\end{table}

%\clearpage
\section{Discussion}

In this paper, we have formulated a series of generalizations of
the CSD analysis method applicable to extracellular recordings in
brain tissue.  This generalization is based on a general theory
that we derived and which aims at linking the extracellular
potential with current source densities in the tissue.  We have
considered a mean-field version of Maxwell equations by considering
the different fields as averages over some reference volume.  By
varying the size of this volume, one can apply the same theory to
different scales.  At microscopic scales ($\sim 1~\mu$m and
smaller), the theory must use the microscopic values for electric
parameters (for example, the very different resistivities of fluids
or membranes).  For mesoscopic or macroscopic scales ($\sim
50~\mu$m and larger), the theory can directly include the
``macroscopic'' measurements of conductivity and permittivity, as
well as their possible frequency dependence if needed.  Note that
this mean-field approach takes into account the physical and
biological properties of the sources, and thus is more general than
previous approaches \cite{BedDes2009a} which did not consider
source densities.  

We have examined different limit cases, such as a purely resistive
extracellular medium with current sources consisting exclusively of
dipoles, in which case the theory recovers the standard model.  In
this standard model, the mean-field theory shows that the electric
potential must be solution of Laplace equation, such that the
``classic'' CSD approach of Mitzdorf~\cite{Mitzdorf} does not
apply. To inverse the CSD in this model, one must apply the forward
solutions of Laplace equation because the associated operator is
non-invertible (see \cite{Pettersen}). In resistive media, the
extracellular potential must have an additional frequency
dependence of \b{$1/f^2$} relatively to that of the current. 
Interestingly, we found that Laplace equation remains valid for
extracellular media which are quasi-resistive (where the electric
parameters weakly depend on frequency).  In this case, the
frequency dependence of the extracellular potential is similar to
that of the current. A weak frequency dependence was indeed found
in some experimental measurements of
resistivity~\cite{Logo2007,Ranck63}, while other
experiments~\cite{Gabriel1996} displayed a much more pronounced
frequency dependence.  With respect to the attenuation with
distance, the standard model predicts an attenuation law as
\b{$1/r^2$}, for both resistive or quasi-resistive media.  

{We also examined the case of monopolar sources.}  If such
monopolar sources are present in addition to dipolar sources,
within resistive or quasi-resistive media, then the CSD equation
takes a slightly different form predicting that the potential will
attenuate asymptotically with the inverse of distance ($1/r$),
while the standard dipole model predicts a square dependence
($1/r^2$).  With monopolar sources, the potential in the
extracellular medium is also solution of Laplace equation, and thus
the inverse algorithm of the CSD method does not apply identically
as for dipoles. To find the inverse CSD, one proceeds similarly as
the standard model by using the ``forward'' solution of Laplace
equation.  However, in this case the sources must be considered as
a linear combination of terms varying as \b{$1/r$} (monopoles) and
\b{$1/r^2$} (dipoles) in this forward solution.  

As a third model, we examined the case of ionic diffusion within
resistive or quasi-resistive media.  In this case, the CSD takes a
form very close to the ``monopolar'' CSD discussed above, but we
found that charge density is frequency dependent according to a
Warburg impedance in \b{$1/\sqrt(\omega)$} {(see
Appendix~\ref{Warburg} and Section \ref{diffusion})}.  This result
is in agreement with a previous modeling study of extracellular
potentials in the presence of ionic diffusion~\cite{BedDes2009a}. 
Another consequence is that, for spherical symmetry, the
attenuation with distance follows a Yukawa potential
\b{$\frac{e^{-k(\omega)r}}{r}$}, which decays faster than the
different laws considered above for large enough frequency (see
Fig.~\ref{appYukawa}). This particular form is responsible for a
low-pass filtering of the extracellular medium.  Note that this
form is obtained for spherical symmetry, but other forms may be
obtained in different geometries.  

{It is important to note that the CSD theory was originally
designed without specific hypotheses about the nature of current
sources~\cite{Plonsey,Nicholson1975,Pitts}, other versions of the
CSD theory clearly assumed that current sources are
dipoles~\cite{Mitzdorf,Pettersen}.  Assuming dipolar sources is
equivalent to assume that we have stationary current conditions at
all scales.  However, we show here (Appendix~\ref{monop}) that at
small scales (synapses), such a stationary current condition is not
necessarily met.} A first possible source of monopolar effects is
the inertia of charge movement along membranes together with
ion-channel selectivity.  Following the opening of ion channels,
the flow of ions will entrain a re-equilibration of the charges
adsorbed on both sides of the membrane.  While this process is
usually considered as instantaneous, together with neglecting
ion-channel selectivity, these processes may have important
consequences.  Indeed, if one takes into account the fact that
charges do not move instantaneously and ion-channel selectivity,
this will necessarily create transient charge accumulation and
monopoles.  A similar effect will occur through ionic diffusion and
electric field, at the interface between the ion channel and the
extracellular medium, because ions diffuse faster than their mean
passage time through the channel, which will also create charge
accumulation and monopolar effects. Note that when this mechanism
produces an external electric field which will contribute to the
extracellular field (in addition to transmembrane currents). These
effects will contribute to transient monopoles, during which
Kirchhoff's node law will not apply. The fact that ions move
considerably slower than electrons in a metal conductor will also
participate to deviations from Kirchhoff laws.  Whether this
transient time is significant, and whether the system could be
continuously ``outside of equilibrium'' due to sustained synaptic
activity, should be investigated by future work.

In the diffusion model, one can directly use the Laplace operator
to inverse the CSD, contrary to the other models.  Taking into
account ionic diffusion requires to revise the ``forward''
approach, because the attenuation law does not follow a \b{$1/r^2$}
law, but rather a Yukawa-type law while the extracellular medium is
associated to a Warburg-type impedance. In a previous study, we
showed that indeed a Warburg type impedance could account for the
transfer function between intracellular and extracellular
potentials~\cite{BedDes2010} (for frequencies comprised between 3
and 300~Hz).  It is also consistent with measurements of
conductivity and permittivity~\cite{Gabriel1996} (but see
\cite{Logo2007}). Note that the linear approximation in the
diffusion model is not valid for high values of the potential
(larger than $\sim$50~mV; see~\ref{Warburg}), so this model applies
well to subthreshold activity, but may need to be revised for
action potentials.  {Similarly, corrections to the CSD given
by the ``forward'' approach (see for example~\cite{Leski2010}) may
also need to be reformulated for non-resistive media.}

Thus, with respect to the paradox of the CSD method, as described
in the introduction, our study suggests that it is naturally solved
by taking into account ionic diffusion. This introduces an
additional source term in the general equation for the electric
potential (see Eq.~\ref{CCC}).  This additional term gives a
Poisson type equation for the potential (instead of Laplace
equation), similar to the classic CSD approach.  Contrary to the
cases with resistive and quasi-resistive media, the classic
algorithm of CSD inversion given by Mitzdorf \cite{Mitzdorf} is
applicable here.  Thus, the results obtained with the classic CSD
analysis are perfectly consistent with ionic diffusion because
diffusion gives a source term which is very close to the
phenomenological model of current source density introduced by
Pitts and Mitzdorf~\cite{Mitzdorf,Pitts}, but in a manner
consistent with Maxwell-Gauss law.  So, we conclude that the usual
approach for CSD inversion, although paradoxical, should
nevertheless give results equivalent to a model with ionic
diffusion and consistent with Maxwell-Gauss law.  

Finally, the few limit cases considered here are by no means
exhaustive.  For example, we neglected the Maxwell-Wagner time of
the extracellular medium and the microscopic variations of
impedance. The theory outlined here is general enough to include
these effects if needed, which is another way to solve the paradox.
For instance, considering phenomena such as ``reactive''
extracellular media, which react to the electric field (for example
through polarization of cell membranes), can be done by taking into
account the Maxwell-Wagner time of the medium (see details in
\cite{BedDes2009a,Bed2006b}). According to Gabriel et
al.~\cite{Gabriel1996}, the macroscopic electric permittivity
becomes larger while macroscopic conductivity becomes smaller for
smaller frequencies, when the electric field is imposed according
to a well-defined direction. In these measurements,
\b{$\omega\tau_{\mbox{\tiny MW}} << 1$} for frequencies larger than
10~Hz, but \b{$\omega\tau_{\mbox{\tiny MW}}$} may be considerably
larger for lower frequencies~\cite{BedDes2009a}, where electric
polarization may play an important role.  The second term in the
lefthand side of Eq.~\ref{CCC} would then not be negligible
anymore. Because this term can be considered as an additional
source term (see \cite{Bed2004}), similar to the case of diffusion,
this also solves the paradox described in the introduction.

In conclusion, we have provided here a generalized CSD approach
valid for more realistic properties of the extracellular medium,
taking into account ionic diffusion or polarization effects,
usually neglected in the standard CSD analysis
\cite{Mitzdorf,Pettersen}.  We found that including such effects
may have deep consequences on the expression to be used for
estimating current sources, and thus may also have consequences on
the values of current sources estimated from experimental
recordings. For example, the potential due to monopolar sources
will decay slower than for dipoles, which will necessarily affect
the recorded potential at the electrode.  Similarly, considering
``reactive'' aspects of the extracellular medium by including a
significant Maxwell-Wagner time leads to a different CSD
expression, close to the form derived for ionic diffusion.  Future
work should apply these expressions to extracellular recordings in
brain tissue, with the aim of identifying which of these phenomena
are most consistent with experimental data.

% ---------------------------------------------------------
% Appendix
% ---------------------------------------------------------

%\clearpage

\begin{appendix}

\section{Appendix}

\subsection{Impedance for systems with ionic diffusion}
\label{activesec}

In this appendix, we consider ionic diffusion at the interface
between ion channels and the extracellular medium, as well as at
the interface with the cytoplasm.  We use the quasistatic
approximation (in the thermodynamic sense), which implies that the
net charge density must be solution of a parabolic partial
differential equation, as for pure diffusion phenomena.  We will
next consider system in spherical symmetry, in which case the
impedance is equivalent to a Warburg impedance.

\subsubsection{Ionic diffusion under the quasistatic approximation
in the thermodynamic sense}

The current density at microscopic scales obeys the equation:
\b{
\begin{equation}
 <\vec{j}>|_m = -\bar{\sigma}_m\nabla <V>|_m +\bar{D}_m~\nabla <\rho>|_m
\label{cour1b}
\end{equation}}
with (see Eq.~\ref{ohm1})
\b{
\begin{equation}
\bar{\sigma}_m = \lambda_q\tau_cn_v(\vec{x},t)
\label{sigb}
\end{equation}}

Let us assume that the system is in a quasi-static case in the
sense of thermodynamics.  As shown by application of Maxwell
distribution of velocity distribution and the principle of detailed
balance~\cite{Vas1983}, we can deduce the Boltzmann distribution
for a field which varies infinitely slow.  This approximation is
valid here because the drift velocity of ions under an electric
field is much lower than the absolute velocity of ions (which is of
the order of sound velocity).  Within this quasistatic
approximation, we can apply the Boltzmann distribution to obtain
the number of ions per unit volume as a function of time and space:
\b{
$$
\begin{array}{ccc}
n_v(\vec{x},t)&=& n_v^{\infty}~[e^{+\frac{<q>|_m<V>|_m (\vec{x},~t)}{kT}} +
e^{-\frac{<q>|_m<V>|_m (\vec{x},~t)}{kT}}]
\end{array}
$$}
when we assume that \b{$V(\infty)=0$} at infinite distance, where
\b{$n_v^{\infty}$} is the number of ions per unit volume at an
infinite distance from the source (``far distance'') and \b{$<q>$}
is the mean absolute charge. 
\b{$k=1.3806503\times~10^{-23}~J/~^\circ K$} is the Boltzmann
constant and \b{$T$} is the temperature in degrees Kelvin.  It
follows that the net charge density is related to the value of the
electric potential according to:
\b{
\begin{equation}
<\rho>|_m(\vec{r},t)= n_v^{\infty}<q>|_m~[e^{+\frac{<q>|_m<V>|_m (\vec{x},~t)}{kT}} -
e^{-\frac{<q>|_m<V>|_m (\vec{x},~t)}{kT}}]
\label{Bolz1a}
\end{equation}}
where \b{$<\rho>|_m(\vec{r}_1,t)$} is the average net charge
density.  Note that the $-$ sign in the second righthand term comes
from the sign of the charge.  Also note that this relation implies
that the net charge density is zero at an infinite distance, and is
linked to the electric potential by a nonlinear relation\footnote{A
consequence of this nonlinear relation is that the medium will
become nonlinear for high values of the electric potential.}.

Applying the operator \b{$\nabla$} on the net charge density gives:
\b{
\begin{equation}
\nabla<\rho>|_m(\vec{x},t)=\frac{(<q>|_m)^2}{kT}~n_v\nabla <V>|_m(\vec{x},t)=\frac{(<q>|_m)^2}{\lambda_q\tau_ckT}\bar{\sigma}_m\nabla <V>|_m
\label{rel1}
\end{equation}}

Note that no such relation would be possible outside of the
quasistatic approximation in the thermodynamic sense.

% ---------------------------------------------------------

\subsection{Non-stationary aspect of the electric field produced by
membrane conductance variations}
\label{CC}

In this appendix, we show that if the transmembrane current is
non-zero, this necessarily implies a charge variation at the
interior of the compartment according to a non-stationary regime
\b{$$\nabla\cdot\vec{\j}+\frac{\partial \rho}{\partial t}=0$$}  We
define as ``interior'' the domain delimited by the inner surface of
the cell membrane as indicated in Fig.~\ref{app}.  We will show
that for different values of stationary current, the net charge
inside a cable compartment is different from zero, and depends on
the value of the current.  Thus, when the current is variable, the
system is necessarily in a non-stationary regime.

To demonstrate this, suppose that we have a current density
$\vec{j}$ which is time independent and stationary, satisfying
\b{$$\nabla\cdot\vec{j}=0 ~ .$$}  The amount of charge situated in
the interior of the compartment strictly inside the compartment can
be calculated from the integral of the electric field across a
closed surface:
\b{
\begin{equation}
 Q_{int}=\oiint\limits_{\partial D}\varepsilon\vec{E}\cdot\hat{n}~dS=\iint\limits_{channel+axial}\varepsilon_c~\vec{E}\cdot\hat{n}~dS
+\iint\limits_{membrane}\varepsilon_m~\vec{E}\cdot\hat{n}~dS
\end{equation}}
where the integral is made over a Gauss surface which goes through
the middle of the membrane thickness, as indicated in
Fig.~\ref{app}.  The surface also avoids ion channels (by
surrounding them below their inner side). 
\b{$\varepsilon=\varepsilon_c$} and \b{$\varepsilon=\varepsilon_m$}
are the electric permittivity of the cytoplasm and of the membrane,
respectively.  

Taking into account Ohm's differential law and stationary current
condition, one obtains:
\b{
\begin{equation}
 I=\iint\limits_{channel+axial} \vec{j}\cdot\hat{n}~dS
=\iiint\nabla\cdot\vec{j}~dv 
=\iint\limits_{channel+axial}\sigma_c~\vec{E}\cdot\hat{n}~dS
= 0
\end{equation}}
where \b{$\sigma_c$} is the electric conductivity of the cytoplasm.
Because the capacitive impedance is infinite for $f=0$, the surface
integral over the membrane is zero, and we obtain:
\b{
\begin{equation}
Q_{int}=\iint\limits_{membrane}\varepsilon_m\vec{E}\cdot\hat{n}dS
\label{gauss}
\end{equation}}
because
\b{$\varepsilon_c\vec{E}=\frac{\varepsilon_c}{\sigma_c}\vec{j}$}
when electric parameters are independent of space.

The cylindric symmetry  together with the isopotentiality of the surface of the
membrane compartment imply that this integral is non zero and
depends on the value of the transmembrane current $I_m$.  If we
have two different transmembrane currents, then the values of the
electric field inside the membrane are necessarily different and
the values of \b{$Q_{int}$} are different too.   In particular, if the
current \b{$I_m$} is zero, this integral equals the negative charge
that there can be inside the compartment at rest.  

It follows that, if we have a variable transmembrane current, the
interior charge varies to satisfy \b{$\nabla\cdot\vec{j} =
-\frac{\partial \rho}{\partial t} \neq 0 $} because the membrane
time constant $\tau_m$ is not negligible.  This shows that the
variation of current sources caused by membrane conductance
variations have the properties of a non-stationary regime.  Note
that this non-stationarity requires to take into account the volume
of the membrane compartment, and would not be present for point
processes.

% ---------------------------------------------------------

\subsection{Ohm's law and frequency dependence}
\label{appenB}

In this appendix, we show that Ohm's law implies that the ratio
between potential and current does not depend on frequency for
frequencies smaller than \b{$10^4~Hz$}.  The aim of the appendix is
to show the physical bases of the model shown in
Section~\ref{diffusion}.

Let us apply an electric field \b{$\vec{E}$}, independent of time
and space, at time \b{$t=0$} and during a time interval \b{$\Delta
T$}, to a homogeneous aqueous solution (similar to salted water) at
thermodynamic equilibrium.  The field will accelerate every ion
according to the following velocity law: 
\newb{
\begin{equation}
 \vec{v}(t_f) = \vec{v}(t_i) + \frac{q_k}{m_q}\vec{E}~ (t_f-t_i)
\end{equation}}
where \newb{$t_i$} is the initial ion collision time and
\newb{$t_f$} its final collision time.
If the field is applied during a time interval \b{$\Delta T$} much
longer than the typical collision time between two molecules (or
ions), then the mean velocity of ions is given by:
\newb{
\begin{equation}
 <\vec{v}(t)>_{\Delta T} = <\vec{v}(t_i)>_{\Delta T}+ \frac{q_k}{m_q^k}\vec{E}<t_f-t_I>_{\Delta T} =
<\vec{v}(t_i)>_{\Delta T} + \frac{q_k}{m_q^k}\vec{E}\tau_c^k 
\end{equation}}
where \b{$\tau_c^k$} the mean collision time of ion $k$, \b{$q_k$}
the charge of ion $k$ and \b{$m_k$} the mass of ion $k$.  Note that
the mean collision time for ions such as
\b{$Na^+,~K^+,~Cl^-,~Ca^{++}$} in sea water are approximately equal
to \newb{$ 10^{-12}--10^{-14}~s$} for temperatures between
\b{$250~~^\circ K$} and \b{$350~~^\circ K$}~\cite{Vas1983}.

If the system is at thermodynamic equilibrium, we have
\newb{$<v(t_i)>_t=0$}.  It follows that the time average of the
current density is given by:
\newb{
\begin{equation}
 <\vec{\j}> = <\sum_{k=1}^{N}   \frac{q_k^2\tau_c^k}{m_q^k}n_v^k>~\vec{E} 
\end{equation}}
where \b{$n_v^k$} is the number of ions of type $k$ per unit
volume.  Because the quantities \b{$\frac{q_k^2}{m_q^k}$},
\b{$\tau_c^k$} and \b{$n_v^k$} are statistically independent, we
can write: 
\newb{
\begin{equation}
 \sigma^e =   \lambda_q\tau_cn_v 
\label{ohm1}
\end{equation}}
where \b{$\lambda_q=<\frac{q_k^2}{m_q^k}>_k$},
\b{$\tau_c=<\tau_c^k>_k$} and \b{$n_v=N<n_v^k>_k$} are average
values of the quantities \b{$\frac{q^2}{m_q}$}, \b{$\tau_c$} and
\b{$n_v$} for each type of ion.  

Thus, this expression shows that the ratio between the time average
of the current and the electric field does not depend on time when
the time interval \b{$\Delta T$} is much longer than the mean
collision time between two molecules of the solution, because the
number of collisions of each ion is very large.  

Because the mean collision time is of the order of
\b{$10^{-12}-10^{-14}~s$} for a temperature included between
\b{$250~~^\circ K$} and \b{$350~~^\circ K$} (see details in
\cite{Vas1983}), we can write that we have a very good
approximation of a variable electric field by a piecewise constant
function (staircase), with time intervals are much longer than the
mean collision times.  Then, the time average of the current
density will be proportional to the applied field.  We can
therefore write that the ratio between mean current density and
applied electric field does not significantly depend on the
frequency of electric field -- nor of time -- for frequencies
smaller than $10^4~Hz$ \footnote{In fact, experiments have shown
that the frequency of the signal much reach the order of
$10^{9}$--$10^{10}$~Hz to evidence a deviation between the
measurements and the formalism given here (see details in De Felice
L.J. 1981. {\it Introduction to Membrane Noise}, Plenum Press, New
York; see also ref.~\cite{Vas1983}).}.  This approximation is
called differential Ohm's law.  

% ---------------------------------------------------------
\subsection{Ionic diffusion in spherical symmetry}
\label{Warburg}

To simplify {the computation of Eq.~\ref{chardens}}, we
consider a model with a spherical source, surrounded by a
spherically-symmetric (or isotropic) extracellular medium with
boundary condition and isopotential over the surface of the source.
In this case, we have the following equality in Fourier space:
\newb{
\begin{equation}
\frac{d^2<\rho_{\omega}>|_m}{dr^2}+\frac{2}{r}
\frac{d<\rho_{\omega}>|_m}{dr} 
+i\frac{\omega}{<\beta>|_m}~<\rho_{\omega}>|_m=0
\label{eqdif}
\end{equation}}
The general solution of this equation is:
\newb{
\begin{equation}
<\rho_{\omega}>|_m(r) = A(\omega)\frac{e^{\sqrt{-i\frac{\omega}{<\beta>|_m}}~r}}{r} + B(\omega)\frac{e^{-\sqrt{-i\frac{\omega}{<\beta>|_m}}~r}}{r}
\end{equation}}
where $r$ is the distance between the geometric center of the
source.  Because we must have no charge accumulation at infinite, 
we must set $A=0$, which gives:
\newb{
\begin{equation}
<\rho_{\omega}>|_m(r) =  <\rho_{\omega}>|_m(R)\frac{Re^{-\sqrt{-i\frac{\omega}{<\beta>|_m}}~(r-R)}}{r}
\label{jjj-1}
\end{equation}}

Taking into account Eq.~\ref{42}, the current density is given by:
\newb{
\begin{equation}
 <j_{\omega}>|_m(r)= <\beta>|_m\frac{\partial <\rho_{\omega}>|_m}{\partial r}  
=<\beta>|_m~(~\frac{1}{r}+\sqrt{-i\frac{\omega}{<\beta>|_m}~}~)~
<\rho_{\omega}>|_m(r)~
\label{jjj}
\end{equation}}

By developing the net charge density (see Eq.~\ref{Bolz1a}) in Taylor
series, we have at first order:
\newb{
\begin{equation}
<\rho>|_m(\vec{x},t)= n_v^{\infty}~<q>|_m~[e^{+\frac{<q>|_m<V>|_m (\vec{x},~t)}{kT}}
-e^{-\frac{<q>|_m<V>|_m (\vec{x},~t)}{kT}}]\approx
2\frac{(<q>_m)^2~n_v^{\infty}}{kT}<V>|_m(\vec{x},t)
\label{Bolz1}
\end{equation}}

At physiological temperature (\b{$310~^o K$}), we can write that
the precision of this linear approximation is larger than
\b{$90~\%$} if the potential is smaller than \b{$15~ mV$} (these
estimates are obtained by replacing Boltzmann constant by the
values of the mean charge, $\approx 2\times 10^{-19}~C$). 
However, the precision drops to about
\b{$8~\%$} for \b{$100 ~mV$}, in which case one must consider up to
the third-order term (\b{$\sim <V>|_m^3$}) in the Taylor expansion.
This is the case for action potentials, and thus the propagation of
the field will become more complex for spikes.

It follows that the impedance between infinite distance and a given
point $P$ at distance $r$ from the source is given by:
\newb{
\begin{equation}
 Z_{\omega}(r)=2\frac{kT}{(<q>|_m)^2<\beta>|_m~n_v^{\infty}}\frac{1}{(\frac{1}{r}+\sqrt{-i\frac{\omega}{<\beta>|_m}})}
\end{equation}}

Thus, the impedance will tend to a Warburg impedance for large
distances from the sources, and for high frequencies.  Moreover, if
the curvature radius is very large, or for planar membranes, one
can set \newb{$R=\infty$}, which gives:
\newb{
\begin{equation}
 Z_f \approx 2\frac{kT}{(<q>|_m)^2<\beta>|_m~ n_v^{\infty}}
\frac{1}{(\sqrt{-i\frac{\omega}{<\beta>|_m}})}=\frac{C}{\sqrt{\omega}}
\end{equation}}
where $C\sim(1-i)$ and thus, the phase of the impedance becomes
independent of frequency.

One sees that the value of the parameter \b{$<\beta>|_m$} and the
curvature radius determine the magnitude of the phase.  They also
determine the distance at which the impedance becomes equivalent to
a Warburg impedance.  For example, the estimates of impedance in
rat cortex made previously~\cite{BedDes2010} indicate that the
impedance can be well approximated by a Warburg impedance for
frequencies above 3~Hz.

\subsection{Kramer-Kronig relations}
\label{kk}

In this appendix, we show that the Maxwell-Wagner time
\b{$\tau_{\mbox{\tiny MW}} =
\frac{\varepsilon_{\omega}}{\sigma_{\omega}^{~e}} \approx
\frac{k}{\omega}$} when the conductivity is very low at zero
frequency, based on Kramers-Kronig relations (see Eq. \ref{kronig}).

These relations show that if \b{$\sigma_{\omega}^{~e}$} does not
depend on frequency, then \b{$\varepsilon_{\omega}$} will also be
frequency independent.  These relations also show that if
\b{$\varepsilon_{\omega}-\varepsilon_{\infty}=k\omega^{-(1-b)}$},
then we have (see Eq.~\ref{kronig})
\b{
$$
\sigma_{\omega}^{~e} -\sigma_0^{~e} = \frac{-2\omega^2}{\pi}\fint_{0}^{\infty} \frac{~~~~k(\omega')^{-(1-b)}}{\omega'^2-\omega^2}~d\omega'
$$}
By setting \b{$x=\frac{\omega'}{\omega}$}, we obtain
\b{
$$
\sigma_{\omega}^{~e} -\sigma_0^{~e}
 =\Big[\frac{-2}{\pi}\fint_{0}^{\infty}
 \frac{1}{x^{1-b}(x^2-1)}dx\Big]~k\omega^{b}=k_1k\omega^b
$$}
where we have by definition
\b{
$$
k_1=\fint_{0}^{\infty}
 \frac{1}{x^{1-b}(x^2-1)}dx
=\lim\limits_{|\epsilon|\to 0}\Big[\int_{0}^{1-|\epsilon|}
 \frac{1}{x^{1-b}(x^2-1)}dx + \int_{1+|\epsilon|}^{\infty}
 \frac{1}{x'^{1-b}(x'^2-1)}dx'\Big]
$$}
By replacing \b{$x=1/x'$} in the second integral in the righthand
term, we get
\b{
$$
k_1
=\lim\limits_{|\epsilon|\to 0}\int_{0}^{1-|\epsilon|}
[x^{-(1-b)}+x^{(1-b)}]\frac{1}{(x^2-1)}dx
$$}

This relation shows that we have
\b{$\frac{\varepsilon_{\omega}-\varepsilon_{\infty}}
{\sigma_{\omega'}^{~e}-\sigma_0^{~e}}=\frac{1}{k_1\omega}=
\frac{\kappa}{\omega}$} (we have set \b{$\kappa=\frac{1}{k_1}$})
such that a small variation of \b{$\sigma_{\omega}$} relative to
frequency (for low frequencies) will entrain a strong variation of
\b{$\varepsilon_{\omega}$} relative to frequency, and thus a very
large value of \b{$\tau_{\mbox{\tiny MW}}$} for low frequencies. 
If \b{$\sigma_0^{~e}\approx 0$} and if \b{$\varepsilon_{\omega}$}
is much larger than that of vacuum, then \b{$\tau_{\mbox{\tiny
MW}}$} is given by:
\b{
\begin{equation}
 \tau_{\mbox{\tiny MW}} = \frac{\varepsilon_{\omega}}{\sigma_{\omega}^{~e}}
\approx \frac{\kappa}{\omega}.
\end{equation}}
where \b{$\kappa$} is a constant which depends on the exponent $b$
(see Fig.~\ref{Dia1}).  This approximation corresponds well to the
experimental measurements of \textit{Gabriel et al} (see
\cite{Gabriel1996}).

Finally, if the conductivity varies very slowly with respect to
frequency (small value of $a$), then the permittivity will be
proportional to \b{$1/\omega$}, and will therefore vary very
steeply at low frequencies.

\end{appendix}

\subsection*{Acknowledgments}

Research supported by the CNRS, Agence Nationale de la Recherche
(ANR Complex-V1, RAAMO), and the European Community (FET grants
FACETS FP6-015879, BrainScales FP7-269921).

% ------------------------------------------------------------
% Biblio
% ---------------------------------------------------------
%\clearage

{
\label{Bibliographie-Debut}

\label{Bibliographie-Fin}
}

% ------------------------------------------------------------
% Figure legends
% --------------------------------------------------------
\clearpage
\normalsize
\section*{Figure legends}

\begin{figure}[h!]

\caption{{(color online)} Attenuation profile of the
extracellular potential as a function of distance.  The profile of
the potential with distance is shown for two positive values of
\b{$<\beta>$}: one value comparable to the diffusion coefficients
of \b{$k^+,Na^+,Cl^-$} (left), and another value 100 times smaller
(right).   When \b{$<\beta>$} is comparable to the diffusion
coefficients, the attenuation according to Yukawa potential is
similar to Coulomb's potential and we have a Warburg impedance. 
For smaller values of \b{$<\beta>$}, the Yukawa potential
determines a significant additional low-pass filter.  The
attenuation law is very steep for frequencies larger than 100~Hz
when \b{$<\beta>|_m = 10^{-11}~m^2/s$}.  The different curves
indicated are the Coulomb's attenuation law as \b{$1/r$}
(monopol<es; solid line) and as \b{$1/r^2$} (dipoles; dotted line;
thick gray lines correspond to attenuation according to a Yukawa
potential; thin lines correspond to Yukawa attenuation combined
with a Warburg impedance, which is the most complete case taking
into account ionic diffusion effects.  In each case, different
frequencies are compared (1, 10, 100 and 1000~Hz) and are shown by
different colors {and dashed lines}, as indicated.  The
current source has a radius of \b{$5~n m$}.}

\label{appYukawa}
\end{figure}

\begin{figure}[h!]

\caption{{(color online)} Scheme of an isopotential membrane
compartment with ion channels.  The blue circles indicate ion
channels in the membrane.  The dotted line indicates a Gauss
surface delimiting the interior of the compartment. The values of
electric permittivity $\varepsilon$ are equal to $\varepsilon_m$ in
the membrane and $\varepsilon_c$ in the cytoplasm.   The electric
conductivity $\sigma$ is equal to $\sigma_c$ in the open channels,
and is assumed to be zero in the membrane.}

\label{app}
\end{figure}

\begin{figure}[h!]

\caption{Representation of \b{$\kappa$} as a function of the
exponent $b$.  The values of \b{$\kappa$} were calculated from
expression \b{$\frac{1}{\frac{-2}{\pi}\fint_{0}^{\infty}
\frac{1}{x^{1-b}(x^2-1)}dx}$}, where \b{$\fint_{0}^{\infty}
\frac{1}{x^{1-b}(x^2-1)}dx =lim_{x\to 1}\int_{0}^{x}
[x^{-(1-b)}+x^{(1-b)}]\frac{1}{(x^2-1)}dx $}.}.  

\label{Dia1}
\end{figure}

% ------------------------------------------------------------
% Figures
% --------------------------------------------------------
\clearpage
\section*{Figures}

\begin{figure}[h!]
\centerline{
\includegraphics[width=\columnwidth]{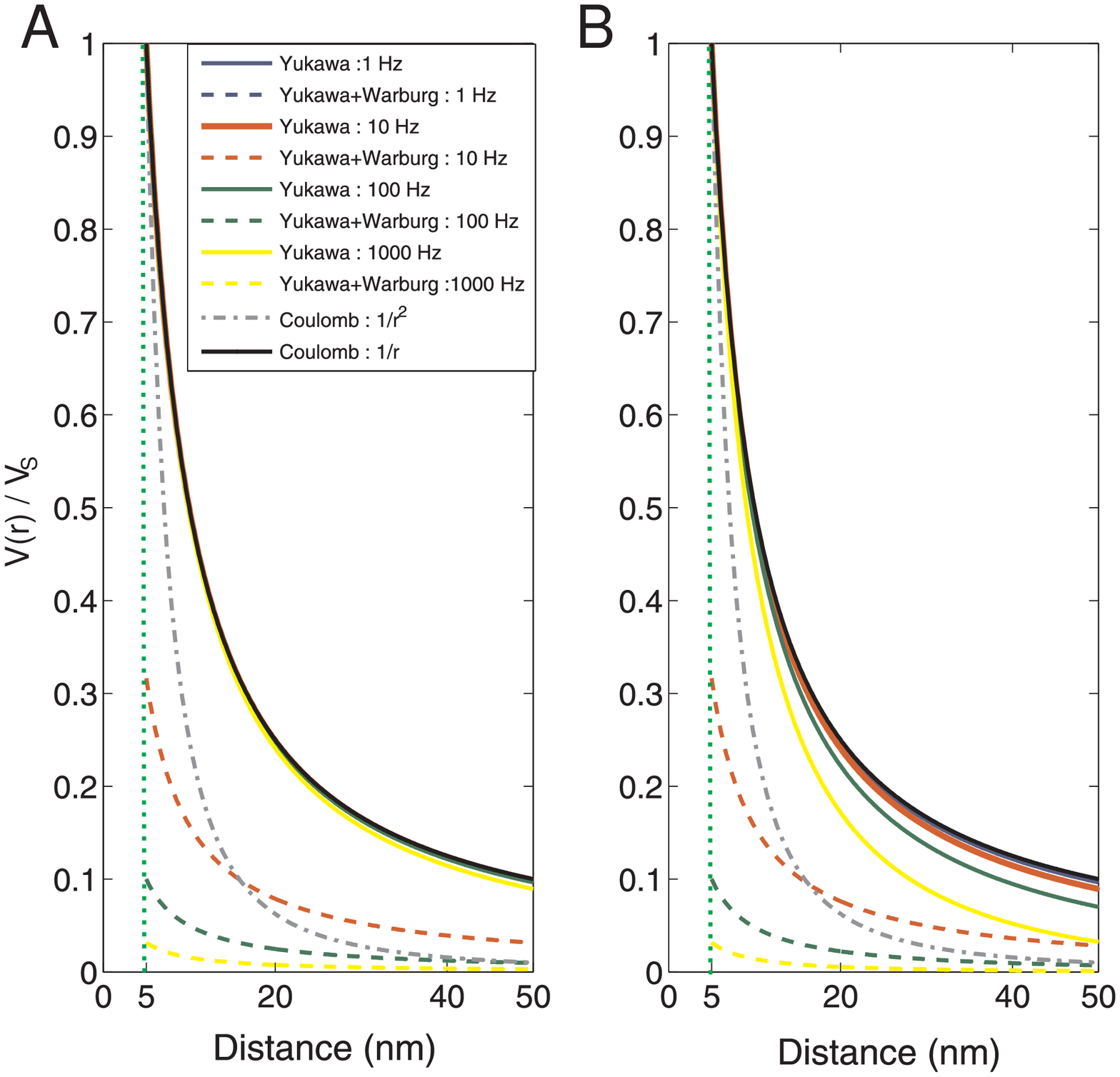}
}

{\bf Figure~\ref{appYukawa}}

\end{figure}

\begin{figure}[h!]
\centerline{
\includegraphics[width=0.5\columnwidth]{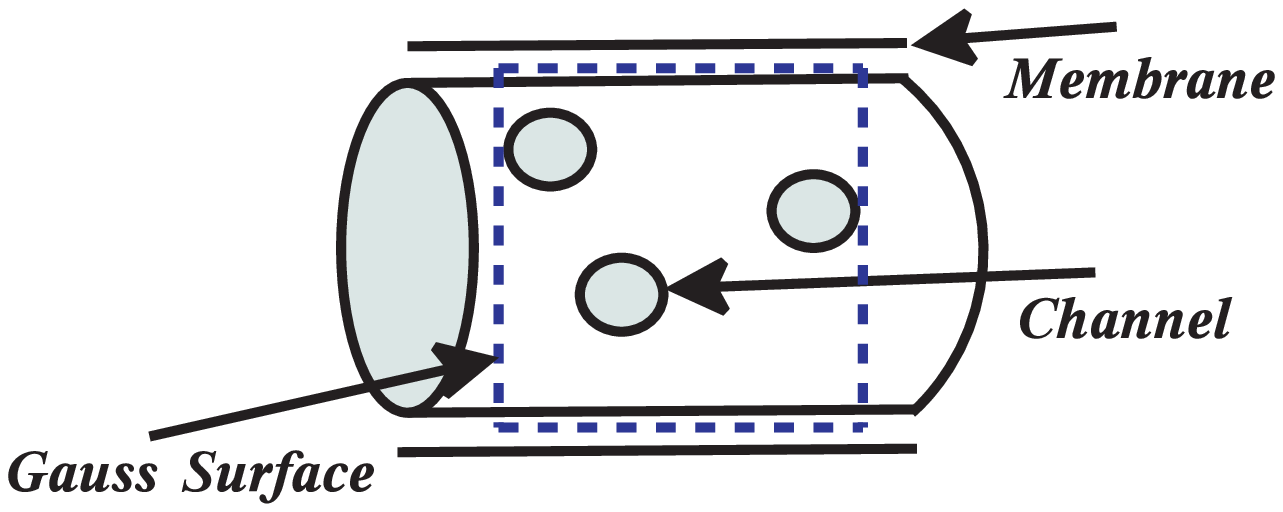}
}

{\bf Figure~\ref{app}}

\end{figure}

\begin{figure}[h!]
\centerline{
\includegraphics[width=0.5\columnwidth]{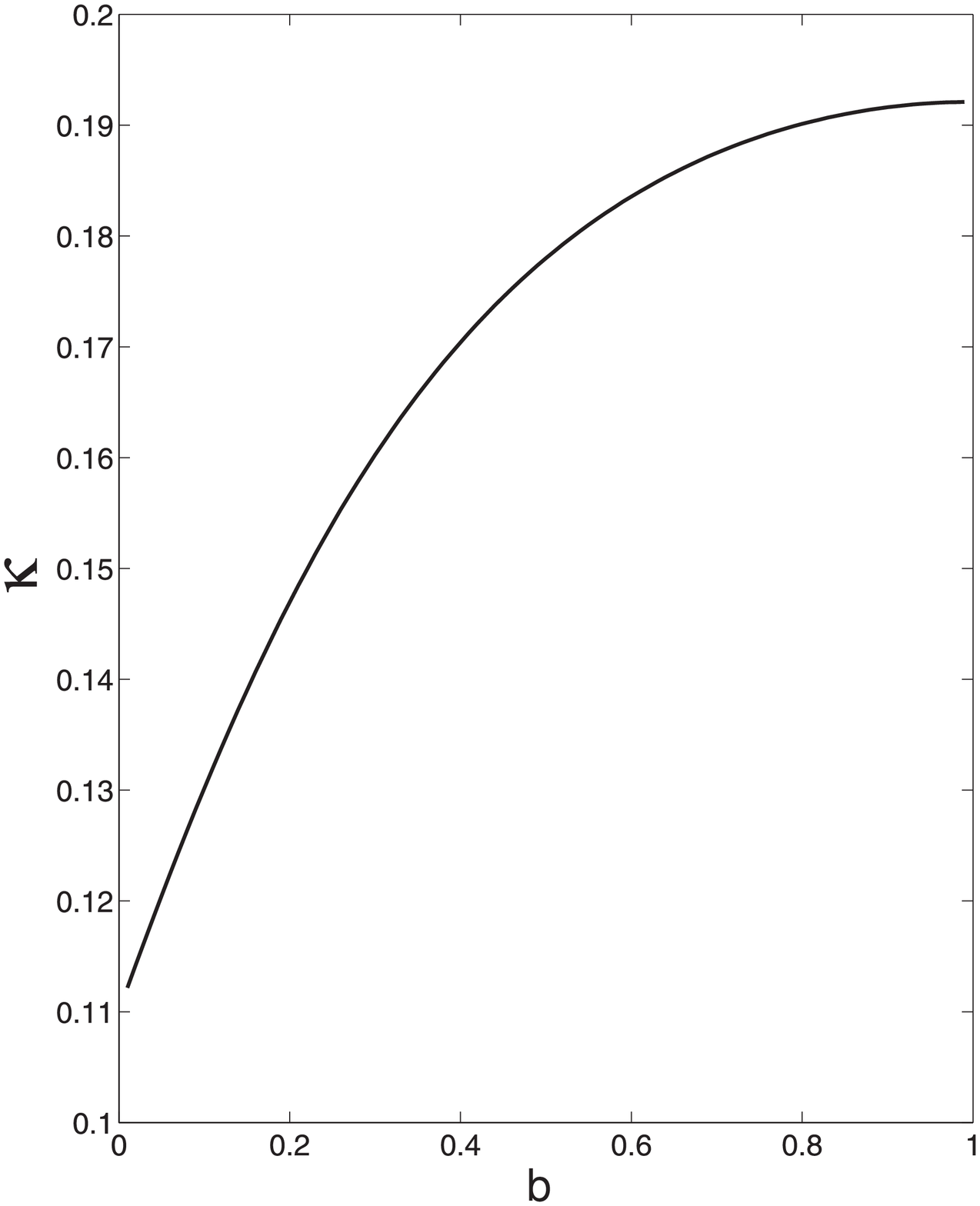}
}

{\bf Figure~\ref{Dia1}}

\end{figure}

\end{document}